 \documentclass{nature}

\usepackage[top=0.8in, bottom=1in, left=0.8in, right=0.8in]{geometry}

\usepackage{graphicx}
\usepackage{amsmath}
\usepackage{amssymb}
\usepackage{threeparttable}
\usepackage{caption}
\usepackage{bibunits}
\usepackage{color}
\usepackage{adjustbox}
\usepackage{url}
\usepackage{soul}
\usepackage[misc]{ifsym} %\Letter
\usepackage{bbding} 
\usepackage{soul}

\usepackage{graphics}
\usepackage{hyperref}
\usepackage[dvipsnames]{xcolor}
\usepackage{newtxtext}
\usepackage[varg,smallerops,upint]{newtxmath}
\usepackage{newfloat}
\DeclareFloatingEnvironment[fileext=lop]{Extended Data Figure}

\newcommand{\fluxunit}{erg\,s$^{-1}$\,cm$^{-2}$}

\renewcommand{\arraystretch}{1.4} % Default value: 1
\defaultbibliography{MasterBiblio.bib}
\defaultbibliographystyle{naturemag}

\title{Accelerated  Formation of Ultra-Massive Galaxies in the First Billion Years}

\author{
M. Xiao$^{1}$, P. A. Oesch$^{1,2}$, D. Elbaz$^{3}$, L. Bing$^{4}$, E. Nelson$^{5}$, A. Weibel$^{1}$, G. D. Illingworth$^{6}$, P. van Dokkum$^{7}$, R. P. Naidu$^{8}$, E. Daddi$^{3}$, R. J. Bouwens$^{9}$, J. Matthee$^{10}$, S. Wuyts$^{11}$, J. Chisholm$^{12}$, G. Brammer$^{2}$, M. Dickinson$^{13}$, B. Magnelli$^{3}$, L. Leroy$^{3}$, D. Schaerer$^{1}$, T. Herard-Demanche$^{9}$, S. Lim$^{14,15}$, 
L. Barrufet$^{1}$, R. M. Endsley$^{12}$, Y. Fudamoto$^{16,17}$, C. Gómez-Guijarro$^{3}$, R. Gottumukkala$^{1}$, I. Labbe$^{18}$, D. Magee$^{6}$, D. Marchesini$^{19}$, M. Maseda$^{20}$, Y. Qin$^{21,22}$, N. Reddy$^{23}$, A. Shapley$^{24}$, I. Shivaei$^{25}$, M. Shuntov$^{2}$, M. Stefanon$^{26,27}$, K. Whitaker$^{28,2}$, J. S. B. Wyithe$^{21,22}$.
\vspace{8pt}}
\begin{document}
\maketitle

\begin{affiliations}
\small
 %1
\item Department of Astronomy, University of Geneva, Chemin Pegasi 51, 1290 Versoix, Switzerland
 %2
\item Cosmic Dawn Center (DAWN), Niels Bohr Institute, University of Copenhagen, Jagtvej 128, K\o benhavn N, DK-2200, Denmark
 %3
\item Universit\'e Paris-Saclay, Universit\'e Paris Cit\'e, CEA, CNRS, AIM, 91191 Gif-sur-Yvette, France
 %4
\item Astronomy Centre, University of Sussex, Falmer, Brighton BN1 9QH, UK
 %5
\item Department for Astrophysical and Planetary Science, University of Colorado, Boulder, CO 80309, USA
 %15
 \item Department of Astronomy and Astrophysics, University of California, Santa Cruz, CA 95064, USA
%12
\item Astronomy Department, Yale University, 52 Hillhouse Ave, New Haven, CT 06511, USA
 %6
\item MIT Kavli Institute for Astrophysics and Space Research, 77 Massachusetts Ave., Cambridge, MA 02139, USA
 %7
\item Leiden Observatory, Leiden University, NL-2300 RA Leiden, Netherlands
 %8
\item Department of Physics, ETH Z{\"u}rich, Wolfgang-Pauli-Strasse 27, Z{\"u}rich, 8093, Switzerland
 %9
\item Department of Physics, University of Bath, Claverton Down, Bath, BA2 7AY, UK
 %10
\item Department of Astronomy, The University of Texas at Austin, 2515 Speedway, Stop C1400, Austin, TX 78712-1205, USA
 %11
\item NSF’s National Optical-Infrared Astronomy Research Laboratory, 950 N. Cherry Ave., Tucson, AZ 85719, USA
 %13
 \item Kavli Institute for Cosmology, University of Cambridge, Madingley Road, Cambridge, CB3 0HA, UK  
 \item Cavendish Laboratory, University of Cambridge, 19 JJ Thomson Avenue, Cambridge, CB3 0HE, UK
\item Waseda Research Institute for Science and Engineering, Faculty of Science and Engineering, Waseda University, 3-4-1 Okubo, Shinjuku, Tokyo 169-8555, Japan
 %14
\item National Astronomical Observatory of Japan, 2-21-1, Osawa, Mitaka, Tokyo, Japan
 %16
\item Centre for Astrophysics and Supercomputing, Swinburne University of Technology, Melbourne, VIC 3122, Australia
 %17
\item Department of Physics and Astronomy, Tufts University, 574 Boston Avenue, Medford, MA 02155, USA
 %18
\item Department of Astronomy, University of Wisconsin-Madison, 475 N. Charter St., Madison, WI 53706 USA
 %19
\item School of Physics, University of Melbourne, Parkville, VIC 3010, Australia
 %20
\item ARC Centre of Excellence for All Sky Astrophysics in 3 Dimensions (ASTRO 3D), Australia
 %21
\item Department of Physics and Astronomy, University of California, Riverside, 900 University Avenue, Riverside, CA 92521, USA
 %22
\item Department of Physics \& Astronomy, University of California, Los Angeles, 430 Portola Plaza, Los Angeles, CA 90095, USA
 %23
\item Steward Observatory, University of Arizona, Tucson, AZ 85721, USA
%24
\item Departament d'Astronomia i Astrof\`isica, Universitat de Val\`encia, C. Dr. Moliner 50, E-46100 Burjassot, Val\`encia,  Spain
%25
\item Unidad Asociada CSIC ``Grupo de Astrof\'isica Extragal\'actica y Cosmolog\'ia" (Instituto de F\'isica de Cantabria - Universitat de Val\`encia)
%26
\item Department of Astronomy, University of Massachusetts, Amherst, MA 01003, USA
\end{affiliations}

\begin{bibunit}

\begin{abstract}

Recent JWST observations have revealed an unexpected abundance of massive galaxy candidates in the early Universe, extending further in redshift and to lower luminosity than what had previously been found by sub-millimeter surveys\cite{Smail1997, Hughes1998, Smail2004, Walter2012, Wang2019, Hodge2020}. These JWST candidates have been interpreted as challenging the $\Lambda$CDM cosmology\cite{Menci2022, Boylan-Kolchin2023, Lovell2023}, but, so far, they have mostly relied only on rest-frame ultraviolet data and lacked spectroscopic confirmation of their redshifts\cite{Naidu2022, Castellano2022, Labbé2023, Pérez-González2023, Finkelstein2023, Willott2024, McLeod2024}. Here we report a systematic study of 36 massive dust-obscured galaxies with spectroscopic redshifts between $z_{\rm spec}=5-9$ from the JWST FRESCO survey. We find no tension with the $\Lambda$CDM model in our sample. However, three ultra-massive galaxies (log$M_{\star}/M_{\odot}$ $\gtrsim11.0$) require an exceptional fraction of 50\,\% of baryons converted into stars -- two to three times higher than even the most efficient galaxies at later epochs. The contribution from an active nucleus is unlikely because of their extended emission. Ultra-massive galaxies account for as much as 17\% of the total cosmic star formation rate density\cite{Madau2014} at $z\sim5-6$.

\end{abstract}

Our sample was obtained by exploiting the imaging and spectroscopic data provided by the JWST FRESCO NIRCam/grism survey\cite{Oesch2023}. Unlike slit spectroscopy, which can only target pre-selected sources, grism spectroscopy can effectively provide a complete sample of emission line galaxies, with the FRESCO survey reaching down to $\sim2\times10^{-18}{\rm erg s}^{-1}{\rm cm}^{-2}$ ($5\sigma$ depth) over a contiguous region of $2\times62$ arcmin$^{2}$ in the GOODS-South and North fields\cite{Giavalisco2004}. Our analysis focuses on all emission line galaxies at spectroscopic redshifts $z_{\rm spec}=5-9$, and includes a detailed study of a subsample of 36 galaxies selected by their red color (F182M $-$ F444W $>$ 1.5 mag; at least one emission line $>8\sigma$; Methods; Extended Data Fig.~\ref{Extended_color_mag}) suggestive of high dust attenuation (typically $A_{\rm V}>1$ mag), hereafter called dusty star-forming galaxies (DSFGs). These dust-obscured galaxies are critical for understanding potential biases in previous studies that may have overlooked such optically invisible yet massive systems\cite{Wang2019, Fudamoto2021, Xiao2023, Barrufet2023, Pérez-González2023b}. With the deep 4-5$\mu$m grism spectra in the F444W filter, we probe H${\alpha}$+[NII] emission lines at $z=4.86-6.69$ and [OIII]$\lambda\lambda4960,5008$+H${\beta}$  at $z=6.69-9.08$. Additionally, the FRESCO survey provides high-resolution NIRCam images (F182M, F210M, and F444W; typical 5$\sigma$ depth of $\sim$28.2 mag), that complement ancillary data from both HST\cite{Whitaker2019} and now also JWST/JADES\cite{Rieke2023}.

\begin{figure*}
	\centering
	\includegraphics[scale=0.23]{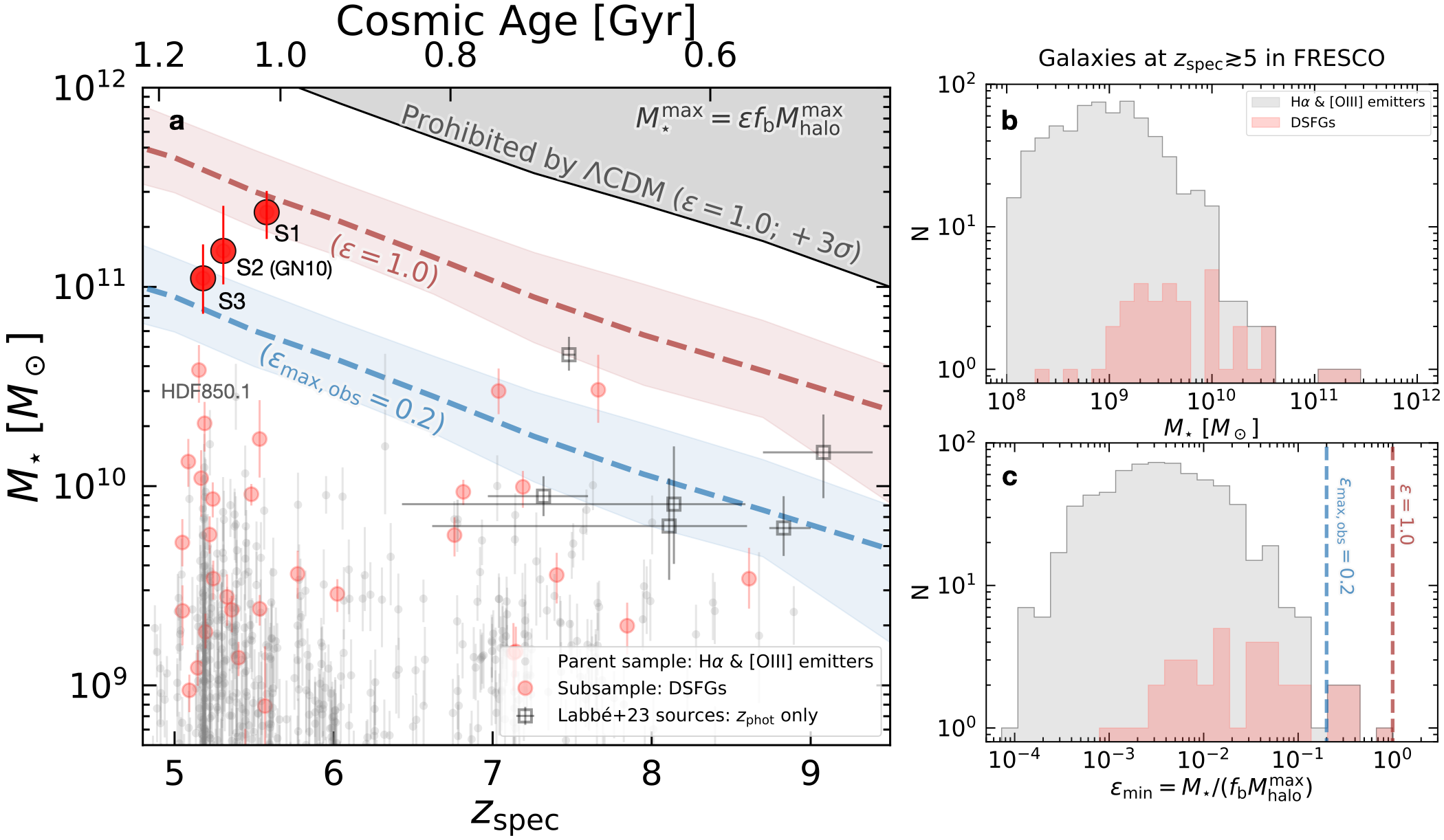} 
	\caption{\small{\textbf{Spectroscopically-confirmed galaxies at $z_{\rm spec}>5$ in the FRESCO fields.} \textbf{a.} Stellar masses of sample galaxies compared to model expectations. The grey-filled circles show the parent sample of H${\alpha}$ \& [OIII] emitters. The red-filled circles show a subsample of 36 red DSFGs. Error bars correspond to stochastic 1$\sigma$ uncertainties from SED fits alone.   The grey-empty squares are massive sources reported in the literature\cite{Labbé2023} that have only $z_{\rm phot}$.  The red and blue dashed lines indicate the maximum stellar mass calculated from the maximum halo mass ($M_\mathrm{halo}^\mathrm{max}$) given the FRESCO survey volume, based on $M_{\star}^\mathrm{max} = \epsilon f_{\rm b} M_\mathrm{halo}^\mathrm{max}$, the cosmic baryon fraction $f_{\rm b} = \Omega_{\rm b}/\Omega_{\rm m} = 0.158$, and assuming a baryon-to-star conversion efficiency of $\epsilon=1$ and 0.2, respectively. Their shaded regions are 1$\sigma$ scatter considering the cosmic variance, based on the FLAMINGO simulations\cite{Schaye2023} (see Methods). Given the uncertainties of cosmic variance, we also show the typical 3$\sigma$ upper limit at $z\sim5.5$ with $\epsilon=1$ as the black line, indicating the stellar mass prohibited by the standard $\Lambda$CDM cosmology\cite{Planck2020}. 
 \textbf{b.} Stellar mass distributions of sample galaxies. 
\textbf{c.} Minimum efficiency distributions of sample galaxies, calculated by $M_{\star}/(f_{\rm b} M_\mathrm{halo}^\mathrm{max})$.
All galaxies have $\epsilon<1$, showing no significant tension with the $\Lambda$CDM model. Three ultra-massive DSFGs (red-filled large circles) have an average value of $\epsilon\sim0.5$, suggesting that their baryons are converted to stars very efficiently. 
}}
	\label{fig1}
\end{figure*}

Thanks to the JWST spectroscopy, we can now obtain more accurate stellar masses by accounting for the true emission line contributions in photometric bands in the SED fit at a fixed $z_{\rm spec}$. The stellar masses ($M_{\star}$) are determined by fitting the UV-to-NIR SED using multiple codes (\texttt{Bagpipes}\cite{Carnall2018}, \texttt{CIGALE}\cite{Boquien2019}) and assuming different star formation history models, which produce consistent stellar mass values within the error range (see Methods for details). The redshift and stellar mass estimates of 36 DSFGs are listed in the Extended Data Table.~\ref{tab1}. Among them, 95\% did not have $z_{\rm spec}$ and $M_{\star}$ measurements before our observations. These 36 galaxies span a redshift range of $z_{\rm spec}=5-9$, with stellar masses covering a wide distribution of log($M_{\rm \star}$/$M_{\odot}$) $=8.3-11.4$ (Fig. ~\ref{fig1}). Overall, the DSFGs ($z_{\rm med,spec}=5.5$ and log($M_{\rm \star, med}$/$M_{\odot}$) $=9.6$) have the same median redshift but are more massive than the parent sample of emission line galaxies ($z_{\rm med,spec}=5.5$ and log($M_{\rm \star, med}$/$M_{\odot}$) $=9.0$). This confirms the high-redshift, dusty, and high-mass nature of this elusive class of galaxies that has so far mainly been studied from photometry alone\cite{Wang2019, Xiao2023, Barrufet2023, Labbé2023, Pérez-González2023b}.

The 36 DSFGs and the parent sample of emission line galaxies are shown in Fig.~\ref{fig1}.  Their stellar masses are compared to the maximum mass at which one would expect to find a galaxy within our survey volume, given the prevailing halo mass function and cosmic baryon fraction\cite{Boylan-Kolchin2023, Lovell2023}. Under this paradigm, we derive the most massive dark matter halo mass (M$_\mathrm{halo}^\mathrm{max}$) at different redshifts within the corresponding FRESCO survey volume ($\sim1.2\times10^{6}$ Mpc$^3$ at $z=5-9$) according to the halo mass function. The maximum stellar mass is inferred from the maximum dark matter halo mass, based on $M_{\star}^\mathrm{max} = \epsilon f_{\rm b} M_\mathrm{halo}^\mathrm{max}$, with a cosmic baryon fraction $f_{\rm b} = \Omega_{\rm b}/\Omega_{\rm m} = 0.158$\cite{Planck2020}, and the maximum theoretical efficiency ($\epsilon$) of converting baryons into stars. Here, we consider two possible cases of $\epsilon$, as shown in Fig.~\ref{fig1}: the highest efficiency from observation-based phenomenological modelings, such as abundance matching and halo occupation distribution models ($\epsilon_{\rm max,obs} = 0.2$; dashed line)\cite{Moster2013, Moster2018, Tacchella2018, Pillepich2018, Wechsler2018, Shuntov2022} and the maximum efficiency logically allowed ($\epsilon=1$; solid line). 

We then compute lower limits on the efficiency of our sample galaxies (Fig.~\ref{fig1}b), to check whether massive high-z galaxies could form stars with unexpectedly high efficiency ($\epsilon > 0.2$ or even $\epsilon > 1$).  In this analysis, we conservatively assume that each galaxy is located in the most massive halo. The minimum efficiency is $\epsilon_{\rm min} = M_{\star} / (f_{\rm b} M_\mathrm{halo}^\mathrm{max}$). In this case, we do not find any galaxy with $\epsilon > 1$, suggesting that our sample does not present significant tension with $\Lambda$CDM. The same conclusion still holds for the subsample of DSFGs, which are representative of a population of very massive dusty galaxies. The reliability of this conclusion comes from the fact that we have both robust stellar masses and redshifts. 

However, we do find some galaxies with $\epsilon > 0.2$, i.e., showing increased efficiency in converting baryons into stars. Three are at $z_{\rm spec}\sim5-6$, and two are at $z_{\rm spec}\sim7-8$. we focus here on the three lower redshift objects since the derived stellar masses are significantly less robust for the $z_{\rm spec}\sim7-8$ objects (see Methods for more details). The three sources at $z_{\rm spec}\sim5-6$ are named S1, S2, and S3. They have been detected by SCUBA2 observations (see method), but only S2 (also known as GN10\cite{Riechers2020}), had reliable redshift and stellar mass measurements before JWST.

\begin{figure*}[h!]
	\centering
	\includegraphics[width=1\textwidth]{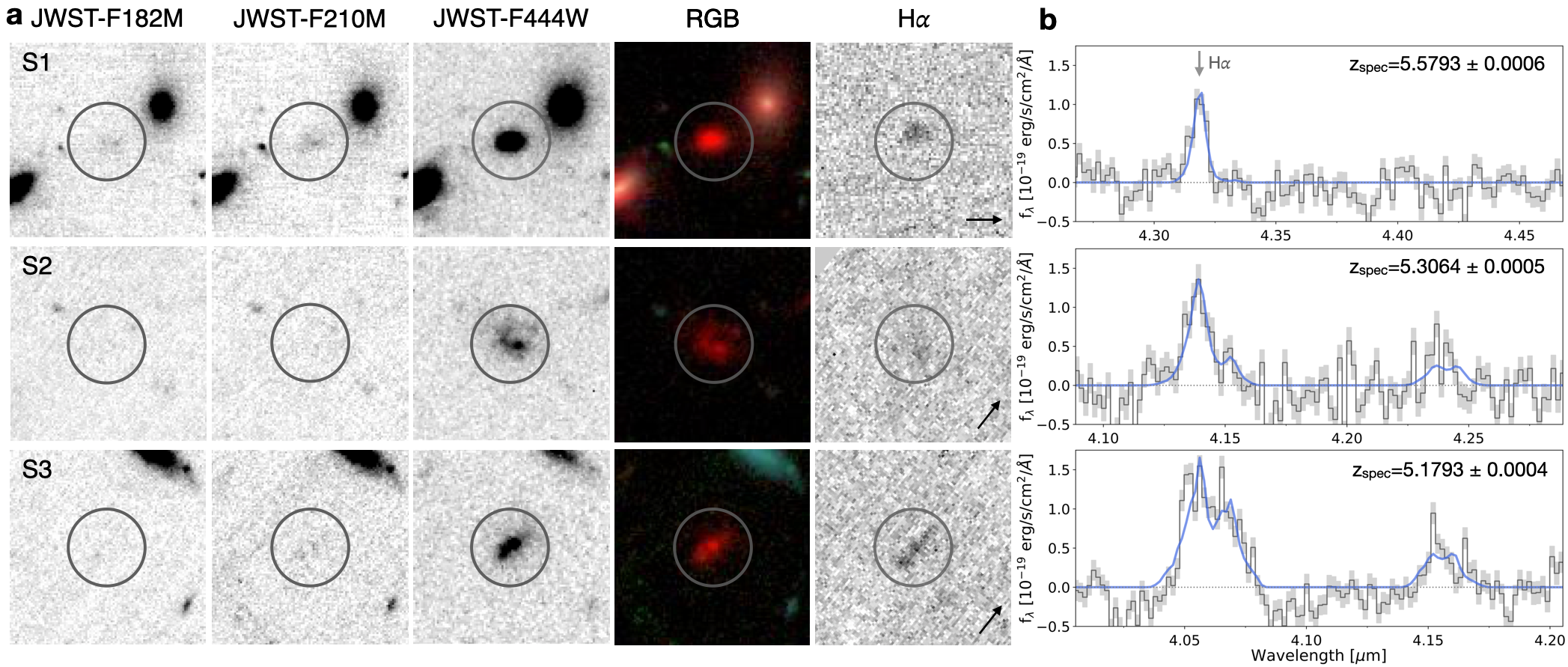}
	\caption{\small{\textbf{Images and spectra of the three ultra-massive galaxies with $\epsilon>0.2$ at $z\sim5-6$. }  
\textbf{a.} 4$^{\prime\prime}$ $\times$ 4$^{\prime\prime}$ stamps obtained in JWST/NIRCam filters (1.82$\mu$m, 2.10$\mu$m, and 4.44$\mu$m), RGB images (F182M in blue, F210M in green, and F444W in red), H$\alpha$ line map. The black arrow in the lower-right corner of the H$\alpha$ line map shows the dispersion direction of the F444W grism. \textbf{b.} 1D spectra (covering H$\alpha$, [NII], and [SII] emission lines) obtained from NIRCam/grism observations with the F444W filter. The gray shaded areas show the associated 1$\sigma$ uncertainty. The best-fit Gaussian line model is shown in blue.}
}
	\label{fig2}
\end{figure*}

Based on the deep JWST observations, we find that S1, S2, and S3 are extremely massive (log($M_{\star}/M_{\odot}$) $\gtrsim11.0$), red (F182M $-$ F444W $>$ 3.5 mag), and heavily dust attenuated ($A_{\rm V}>3$ mag; Figs.~\ref{fig1}~\ref{fig2}, Extended Data Fig.~\ref{Extended_color_mag}, Extended Data Table.~\ref{tab1}).  Self-consistently, we find that they also have high dust-obscured star formation rates (SFR$_{\rm IR}$), which are obtained from far-infrared SED fits with \texttt{CIGALE}, specifically $795\pm40 M_{\odot}$ yr$^{-1}$ for S1, 1,030$_{-150}^{+190}$ M$_{\odot}$ yr$^{-1}$ for S2, and $988\pm49$ M$_{\odot}$ yr$^{-1}$ for S3 (see Methods). This indicates that they are in the process of very efficient stellar mass build-up. We find no signs of a significant contribution to the rest-frame optical light by active galactic nuclei (AGN) based on our investigation of the emission lines, source morphology, and multi-wavelength data. Therefore, we conclude that the ultra-massive nature of these three galaxies is reliable.

By comparing the masses of our galaxies with the predictions in Fig.~\ref{fig1}, it is clear that these sources require an extremely efficient conversion of all available baryons to stars of about 0.5 on average -- two to three times the highest efficiency observed at lower redshift ($\epsilon_{\rm max,obs} \simeq 0.2$)\cite{Moster2013, Moster2018, Tacchella2018, Pillepich2018, Wechsler2018, Shuntov2022}. These three galaxies lie at $z_{\rm spec}\sim5-6$, demonstrating that the existence of ultra-massive galaxies that challenge our galaxy assembly models is not restricted to the most distant Universe at $z>8$\cite{White1978, Boylan-Kolchin2023, Labbé2023}, but includes galaxies at later times that were previously hidden by heavy dust obscuration.

\begin{figure*}
	\centering
	\includegraphics[scale=0.3]{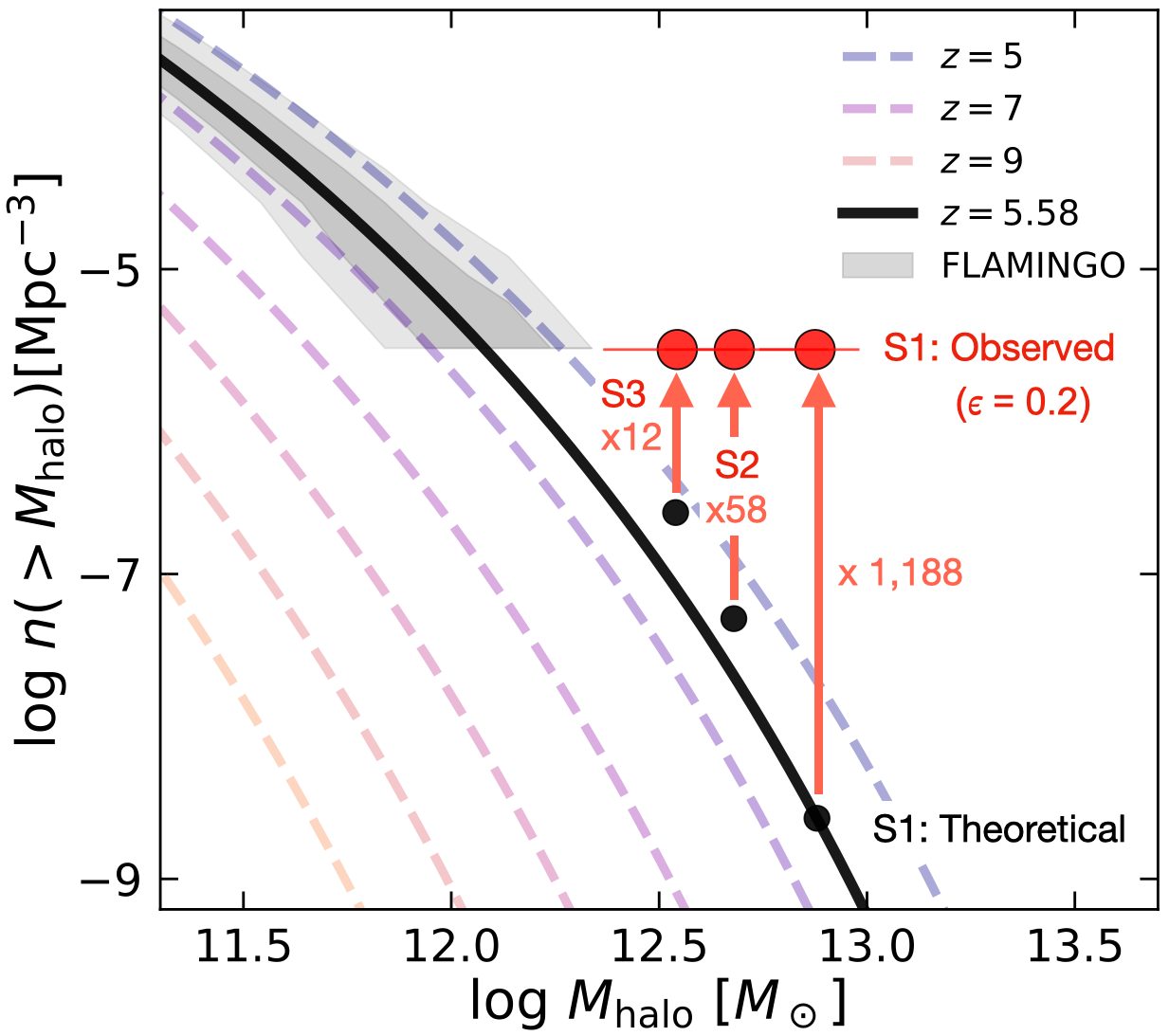}
	\caption{\small{\textbf{Cumulative comoving number density of dark matter halos as a function of halo mass at different redshifts. }  The dark and light grey shaded areas indicate the 1$\sigma$ and 2$\sigma$ scatter of the cosmic variance from the FLAMINGO simulations\cite{Schaye2023}, respectively. We show the halo mass of S1, S2, and S3 assuming a maximum observed efficiency of $\epsilon_{\rm max, obs} = 0.2$. Their observed number densities derived from the FRESCO survey are shown as the red points. Error bars correspond to 1$\sigma$ uncertainties. The theoretical number density of S1, S2, and S3 in the same halo mass from the standard $\Lambda$CDM cosmology (black points) is 1,188$_{-933}^{+2,780}$ times, 58$_{-45}^{+549}$ times, and 12$_{-9}^{+40}$ times lower than the observations, respectively. 
} 
	\label{fig3}}
\end{figure*}

The high required baryon-to-stellar conversion efficiency in the three ultra-massive galaxies could also be demonstrated the other way around. Assuming a maximum observed efficiency of $\epsilon_{\rm max,obs} = 0.2$\cite{Moster2013, Wechsler2018, Shuntov2022}, the stellar masses of S1, S2, and S3 correspond to dark matter halo masses of log($M_{\rm halo}/M_{\odot}$) $=12.88^{+0.11}_{-0.13}$, $12.68^{+0.23}_{-0.17}$, and $12.54^{+0.17}_{-0.18}$, respectively. The observed volume density is $\sim 3.0\times10^{-6}$ Mpc$^{-3}$ at $z\sim5-6$ in the 124 arcmin$^{2}$ of the FRESCO survey fields. For the most extreme case, S1, compared to the theoretical cumulative halo number densities of $\sim 2.8\times10^{-9}$ Mpc$^{-3}$ at log($M_{\rm halo}/M_{\odot})=12.88$, the probability of detecting such a source in a random field as large as the FRESCO survey is only 0.0008$_{-0.0006}^{+0.0031}$ (Fig.~\ref{fig3}). In other words, if the galaxy distribution in the Universe was homogeneous, and if $\epsilon = 0.2$, we would only have expected to detect one source such as S1 in a field 1,188$_{-933}^{+2,780}$ times the area of FRESCO. For S2 and S3, the probabilities are 0.017$_{-0.016}^{+0.060}$ and 0.08$_{-0.06}^{+0.25}$, respectively, which are also low enough to require 58$_{-45}^{+549}$ and 12$_{-9}^{+40}$ times the FRESCO field to detect them, respectively. This shows that the star-formation efficiencies in these galaxies must be significantly higher than normally found at lower redshifts in standard galaxy formation models within cold dark matter halos\cite{White1978, Boylan-Kolchin2023, Dekel2023, Li2023}.

\begin{figure*}
	\centering
	\includegraphics[scale=0.5]{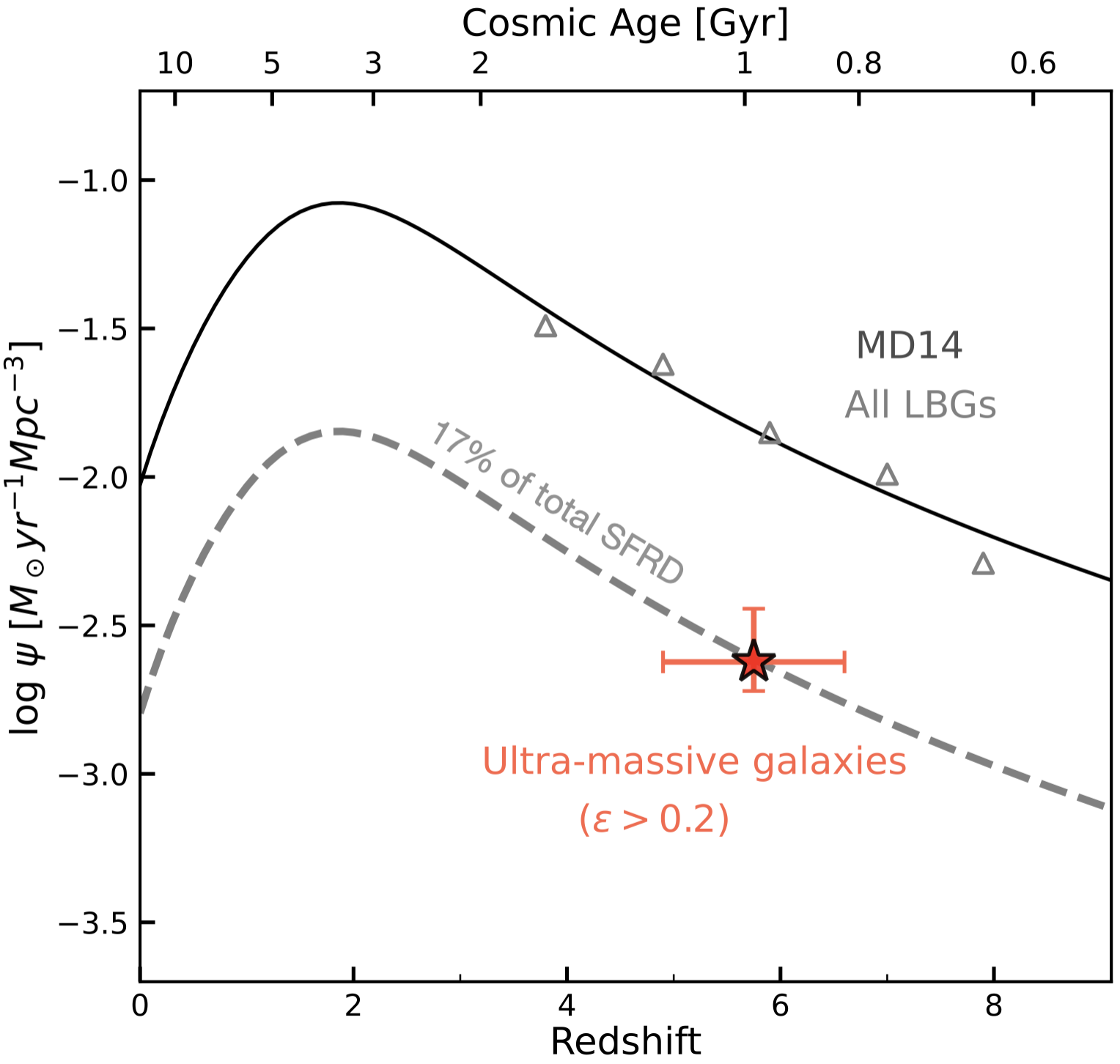}
	\caption{\small{\textbf{Contribution of ultra-massive galaxies to the cosmic SFR density.} 
 The black line is the computation of the total cosmic SFR density (SFRD), $\psi$, corrected for dust attenuation, as a function of redshift by Madau \& Dickinson (MD14)\cite{Madau2014}. At $z\gtrsim4$, the curve is primarily based on UV-selected galaxies (Lyman-break galaxies, i.e., LBGs; grey open triangles)\cite{Bouwens2012a, Bouwens2012b}. 
  The red star corresponds to the SFRD from the spectroscopically-confirmed ultra-massive DSFGs (i.e., S1, S2, and S3) from the FRESCO survey with $\epsilon>0.2$. The error bar on the y-axis corresponds to the 1$\sigma$ uncertainty in $\psi$, while the error bar on the x-axis represents the redshift coverage of our H${\alpha}$ sample ($z=4.9-6.6$).
}
	\label{fig4}}
\end{figure*}

Extremely massive galaxies contribute significantly to the total cosmic star-formation rate density in the early Universe. 
We derive this in the redshift bin given by the H${\alpha}$ sample ($z=4.9-6.6$). Including only S1, S2, and S3, the SFR density reaches 2.4$^{+1.2}_{-0.5}$ $\times$ 10$^{-3}$ $M_{\odot}$ yr$^{-1}$Mpc$^{-3}$ at $z$ $\sim$ 5.8. Compared to the total cosmic SFR density\cite{Madau2014} (Fig.~\ref{fig4}) at $z\sim5.8$, we find that ultra-massive galaxies with efficient baryon-to-star conversion ($\epsilon > 0.2$) alone can account for as much as 17$^{+8}_{-3}$\% of the cosmic SFRD at $z\sim5.8$. This finding suggests a substantial proportion of extremely efficient star formation in the early Universe.

With the secure spectroscopic redshift and stellar mass measurements, our results provide strong evidence that the early Universe has to be two to three times more efficient in forming massive galaxies than the average trend found by previous studies at later times. Our discovery, together with the possible excess of UV-luminous galaxies at $z>8$ revealed by JWST observations, indicates that early galaxy formation models may need to be revised, although within the framework of the $\Lambda$CDM cosmology. We note that two out of the three ultra-massive galaxies have recently been found to be located in a large-scale structure in the process of formation\cite{Herard-Demanche2023}. Therefore, the potential effects of cosmic variance need to be carefully taken into account before designing new models. Our study indicates that the most massive galaxies located in the densest regions of the universe may have a specific formation history, which requires unique models of galaxy formation, such as feedback-free starburst\cite{Dekel2023, Li2023}, to explain how star formation is actually enhanced at a significant rate in these regions. Variations in the IMF could also possibly reproduce the observed extreme properties of our sample. However, these scenarios remain to be investigated by more detailed observations. With higher spatial resolution and/or sensitivity, future ALMA/NOEMA and deep JWST spectroscopic observations could help consolidate the massive nature of these galaxies through dynamical mass measurements, and test different scenarios for their formation with the analysis of the kinematics and chemical composition of the interstellar medium.

\clearpage

\putbib[MasterBiblio]

\end{bibunit}

\clearpage
\appendix

\begin{bibunit}

\noindent\textbf{\large Methods}

\subsection{Cosmology.}
Throughout this paper, we assume a Planck cosmology\cite{Planck2020} and adopt a Chabrier IMF\cite{Chabrier2003} to estimate stellar masses (M$_*$) and star formation rates (SFR). When necessary, data from the literature have been converted with a conversion factor of M$_*$ (Salpeter IMF)\cite{Salpeter1955} = 1.7  $\times$ M$_*$ (Chabrier IMF). All magnitudes are in the AB system\cite{Oke1983}, such that $m_{\rm AB} = 23.9 - 2.5$ $\times$ log(S$_{\nu}$ [$\mu$Jy]).

\subsection{FRESCO NIRCam grism spectra and imaging data.} 

The NIRCam data used in this paper stem from the JWST FRESCO survey\footnote{\url{https://jwst-fresco.astro.unige.ch}} (GO-1895; PI: P. Oesch; see\cite{Oesch2023} for details). FRESCO obtained direct images in three filters (F182M, F210M, and F444W) as well as NIRCam/grism spectroscopy in the F444W filter over $\sim$62 arcmin$^2$ in each GOODS field, North and South, through two 2$\times$4 NIRCam/grism mosaics. The grism spectra span a wavelength of 3.8 to 5.0 $\mu$m, with some parts of the mosaic having slightly reduced coverage. With an exposure time of 7,043 s, the grism data reach an average 5 $\sigma$ line sensitivity of 2$\times$10$^{-18}$ \fluxunit\ at a resolution of R$\sim$1,600. The FRESCO data were acquired between November 2022 and February 2023.

We use the publicly available \texttt{grizli} pipeline\footnote{\url{https://github.com/gbrammer/grizli}} to reduce the slitless NIRCam/grism data (see also Brammer et al, in prep). The raw data are obtained from the MAST archive, and calibrated with the standard JWST pipeline, before being aligned to a Gaia-matched reference frame. 

In order to produce a line-only dataset we remove the continuum along the dispersion direction following\cite{Kashino2022}. We subtract a running median filter with a central 12-pixel gap (corresponding to $\sim20\AA$ rest-frame) along each row of the grism images. To minimize the self-subtraction of bright lines, the filtering is done in two passes, masking pixels with significant flux after the first pass. 

Based on the direct F444W image an individual kernel is created for each source in order to perform optimal extractions of 1D spectra. Slightly modified sensitivity curves and spectral traces are used taking the publicly available v4 grism configuration files\footnote{\url{https://github.com/npirzkal/GRISMCONF}} as the starting point.

In parallel to the long-wavelength grism data, FRESCO obtained short-wavelength imaging in F182M and F210M. At the end of each grism exposure, direct images in F444W are obtained to ensure that the grism data can be aligned. The 5 $\sigma$ depths of the direct images in the F182M, F210M, and F444W filters are 28.3, 28.1, and 28.2 mag, respectively, as measured in circular apertures of 0\farcs32 diameter.

\subsection{Ancillary data and multi-wavelength catalogs.}

In addition to the FRESCO NIRCam data, we also make use of public HST and JWST imaging over the GOODS fields. Most importantly, this includes the HST/ACS and WFC3/IR data from the original GOODS survey\cite{Giavalisco1996} as well as CANDELS\cite{Koekemoer2011, Grogin2011}. For a full list of all HST programs covering this field see the Hubble Legacy Field (HLF) release page\footnote{\url{https://archive.stsci.edu/prepds/hlf/}} (see also\cite{Whitaker2019} and\cite{Illingworth2016}). In GOODS-South, our data cover the Hubble Ultra Deep Field (HUDF\cite{Beckwith2006}) and we include the deep JADES NIRCam imaging that was released in June 2023\cite{Eisenstein2023, Rieke2023}, which incorporates the JEMS NIRCam medium-band imaging\cite{Williams2023}. The images are all co-aligned and drizzled to a common 40 mas/pixel frame. 

Multi-wavelength catalogs are derived using \texttt{SExtractor}\cite{Bertin1996} in dual image mode, using the longest wavelength NIRCam wide filter, F444W, as the detection image. Fluxes are measured in 0\farcs16 radius circular apertures in images that are PSF-matched to the F444W band. Total fluxes are derived from the Kron AUTO aperture provided by the \texttt{SExtractor} in the F444W band, in addition to a correction based on the encircled energy of the Kron aperture on the F444W PSF. 

Photometric redshifts are estimated through the SED-fitting code \texttt{EAZY}\cite{Brammer2008}, using the \texttt{blue\_sfhz\_13} template set which imposes redshift-dependent SFHs, disfavoring SFHs that start earlier than the age of the Universe at a given redshift\footnote{\url{https://github.com/gbrammer/eazy-photoz/tree/master/templates/sfhz}}. We apply an error floor of 5\% prior to running \texttt{eazy} to account for possible remaining systematic uncertainties in the photometric fluxes\cite{Labbé2023} and to allow for more flexibility in the SED-fitting. The methods used to derive the multi-wavelength catalogs with photometric redshifts are presented in more detail in ref.\cite{Weibel2024}.
\\
\\
\textbf{Far-infrared and millimeter data.} In the GOODS-South field, the FRESCO survey overlaps with about 80\% of the GOODS-ALMA 1.1mm survey\cite{Franco2018, Gómez-Guijarro2022, Xiao2023} (PI: D. Elbaz).  The GOODS-ALMA survey covers a continuous area of 72.42 arcmin$^2$ and has a combined dataset\cite{Xiao2023} of high- and low-resolution 1.1mm observations obtained from ALMA Cycle 3 and Cycle 5. The combined map achieves an rms sensitivity of $\sigma$ $\simeq$ 68.4\,$\mu$Jy beam$^{-1}$ with a spatial resolution of 0\farcs447 $\times$ 0\farcs418. The GOODS-ALMA 2.0 source catalog is presented in ref.\cite{Gómez-Guijarro2022}. In addition, the field is also covered by the JCMT/SCUBA-2 at 850$\mu$m\cite{Cowie2018}. The source S1, from our sample in GOODS-S, also has been observed by ALMA band-7 observations as a follow-up to a JCMT/SCUBA-2 target\cite{Cowie2018}. In the GOODS-North field, the FRESCO coverage overlaps with the 450$\mu$m and 850$\mu$m surveys of JCMT/SCUBA-2\cite{Cowie2017, Barger2022}. 
\\
\\
\subsection{Dusty star-forming galaxies (DSFGs) at $z_{\rm spec}\gtrsim5$.}
The traditional approach to systematically selecting DSFGs in the optical to near-infrared wavelengths is to select optically dark/faint galaxies, which are usually based on photometric data only (pre- and present-JWST era). They are required to be faint or completely undetected at optical wavelengths, but bright in the infrared (e.g., $H>$ 27 mag \& [4.5] $<$ 24 mag for $H$-dropouts\cite{Wang2019}; $H>$ 26.5 mag \& [4.5] $<$ 25 mag for optically dark/faint galaxies\cite{Xiao2023}; see also\cite{Franco2018, Alcalde2019, Williams2019, Barrufet2023, Gómez-Guijarro2023, McKinney2023, Akins2023, Pérez-González2023b, Barro2024, vanderVlugt2023}). These magnitude and/or color cuts are designed to select red galaxies with strong Balmer or 4,000$\AA$ breaks at $z\gtrsim3$, so they should be either quiescent/passive galaxies or dusty star-forming galaxies with significant dust attenuation.  The $H$-band magnitude cut helps to avoid significant contamination from passive galaxies and low-$z$ sources\cite{Xiao2023}, but obviously, they cannot be entirely ruled out.

Now, by taking advantage of the imaging and spectroscopic data provided by the JWST FRESCO survey, we no longer need to use the magnitude cuts described above to select optically dark/faint galaxies. The FRESCO grism spectra help us more securely identify high-$z$ star-forming sources through strong nebular emission lines. The spectral coverage of the FRESCO survey with F444W filter allows detections of either H${\alpha}$+[NII]+[SII] lines or [OIII]$\lambda\lambda4960,5008$+H${\beta}$ lines at $z=4.9-9.0$ continuously.

We select candidate galaxies that are red (F182M $-$ F444W $>$ 1.5 mag; Extended Data Fig.~\ref{Extended_color_mag}) and have strong [OIII]$\lambda\lambda4960,5008$+H${\beta}$ or H${\alpha}$+[NII]+[SII] emission line detections. The color threshold we set is similar to the lower color limit of optically dark/faint galaxy selection criteria in the literature\cite{Xiao2023, Gómez-Guijarro2023}, which helps to further compare our sample with those of the pre-JWST/spectra sample. We require all the galaxies to have $>$5$\sigma$ detections at F444W. With some galaxies not detected at F182M, we use the 2$\sigma$ limit as an upper limit in the color selection. For the emission line detection, we require the strongest line to be $>8\sigma$ and the second strongest line to be $>3\sigma$. For some galaxies with only one line detected, we require that the line has $>8\sigma$ detection and their $z_{\rm spec}$ must be within the 16-84th percentile uncertainties of the $z_{\rm phot}$ from the UV-to-NIR SED fitting with \texttt{EAZY} code.  Finally, we have 36 galaxies at $z_{\rm spec}\gtrsim5$ in our sample. Among these, 22 out of 36 galaxies have at least two emission line detections or $z_{\rm spec}$ measurements confirmed from the literature (see Extended Data Table.~\ref{tab1}).

The Extended Data Fig.~\ref{Extended_color_mag} presents our sample selection results on the full FRESCO field, as well as the comparison with previous selection methods and model expectations. Compared to previous selections of $H$-dropouts from Spitzer/IRAC\cite{Wang2019}, our method extends the identification of optically dark/faint galaxies to fainter F444W/4.5$\mu$m fluxes and bluer colors. 

In Extended data Fig.~\ref{Extended_color_mag}, we present a comparison between the color and magnitude of our sample and the model-predicted SED of galaxies. The redshift evolution track of galaxies across the color-magnitude diagram is computed using BC03\cite{Bruzual2003} stellar population synthesis model with constant star formation history starting at $z=10$. The SEDs are further reddened using a Calzetti dust extinction law with E(B-V) at 0.2, 0.4, 0.6, and 0.8 to simulate the color and magnitude evolution of dusty star-forming galaxies. The evolution tracks of galaxies with a total stellar mass of $10^{10} M_\odot$ and $10^{11} M_\odot$ are presented.

While the optically dark/faint selection criteria are defined to target dust-obscured sources, galaxies can also exhibit red F182M $-$ F444W colors due to emission line contamination in the F444W filter. Thanks to the FRESCO emission line measurements, we can now correct this and check which sources only appear red due to line emission. Therefore, we subtract the measured line fluxes from the broad-band photometry to derive pure continuum flux measurements (F444W$^c$). This shows that a small number ($\sim$10 sources) at the faint end only satisfies our selection criteria due to line boosting (see Fig.~\ref{Extended_color_mag}). The brighter galaxies discussed in the main text, however, are only a little affected by this.

The emission-line corrected colors further ensure a fair comparison between our sample with the evolution tracks. This clearly reveals that our JWST imaging+spectroscopic selection could (1) select DSFGs with high purity and without contamination of passive galaxies, and (2) extend the selection of dusty galaxies to lower $M_{\star}$, lower amount of (but still significant) dust reddening and higher $z_{\rm spec}$. 

Our stellar mass estimation suggests the FRESCO dusty star-forming galaxy sample has log($M_{\rm \star}/M_{\odot}$) $=8.3-11.4$ and $A_V=0.1-3.7$ at $z_{\rm spec}>5$ (see Extended Data Table~\ref{tab1} and the section on stellar mass measurements for further details). For comparison, the traditional methods of selecting dusty galaxies (grey triangular regions\cite{Wang2019} in Extended Data Fig.~\ref{Extended_color_mag}; see also\cite{Xiao2023, Alcalde2019, Barrufet2023, Gómez-Guijarro2023}) miss the majority of high-$z$ and(or) faint sources in our sample, but still exclusively select the three ultra-massive and most reddened galaxies with log($M_{\star}/M_{\odot}$) $\gtrsim11.0$ at $z_{\rm spec}\sim5.5$. This further suggests the power and importance of JWST data in providing a complete picture of dusty galaxies in the early Universe.
\\
\\
\textbf{Stellar masses.}
The physical properties of these 36 DSFGs in our parent sample are estimated by fitting the UV-to-NIR SED from the JWST+HST photometry using \texttt{BAGPIPES}\cite{Carnall2018}, with the $z_{\rm spec}$ fixed. We assume a constant star formation history (SFH), the Bruzual stellar population models\cite{Bruzual2003}, and a Calzetti dust attenuation law\cite{Calzetti2000}. We adopt a broad metallicity grid from 0.1 to 2.0 $Z_{\odot}$, an $A_{\rm V}$ grid from 0 to 5 mag, and an age grid from 10 Myr to 1.5 Gyr. For our SED fits, we use the emission line subtracted F444W fluxes in order to remove one free parameter of the emission line model and only fit the stellar continuum emission. We have also ensured that no other strong emission lines are contaminating other photometric bands. From this, we obtain the main physical properties of the 36 DSFGs: $M_{\star}$, $A_{\rm V}$, and SFR. These properties of our sample are listed in Extended Data Table \ref{tab1}. Their locations in the star-formation main sequence (SFMS) are shown in Extended Data Fig.~\ref{Extended_MS}. We would like to emphasize that due to the lack of far-infrared and millimeter data for most of the sources, the physical properties here are only obtained from UV-to-NIR SED fits. In this case, the $M_{\star}$ and SFRs of these dusty star-forming galaxies (and the SFR densities in Fig.~\ref{fig4}) might be underestimated, as they may contain hidden dust regions that absorb all the UV photons, which cannot be reproduced with dust extinction corrections\cite{Xiao2023, Elbaz2018, Puglisi2017}. However, the possible underestimation of $M_{\star}$, SFR, and therefore the SFR density, would make our main results in this paper even more significant. A further comparison of the SFRs obtained from the UV-to-NIR fit with those derived from the IR SED  for the three ultra-massive galaxies can be found in the section on Infrared luminosity and obscured star formation. Examples of best-fit SEDs for three ultra-massive galaxies are in Extended Data Figs.~\ref{Extended_S1}, \ref{Extended_S2}, and \ref{Extended_S3}.

We note that although we could accurately remove the contamination of the emission line in the F444W broad-band photometry, the contribution of the nebular continuum could still be important at shorter wavelengths. To check the possible impact of the nebular continuum, we performed SED modeling on our sample with the original FRESCO and HST broad-band photometry and stellar+nebular model of \texttt{BAGPIPES}. Here we model the nebular emission with a range of ionization parameters log$U$ from -4 to -2 and set the other parameters to be the same as the line-free modeling described above. With these setups, we obtain stellar mass consistent with that from the line-free modeling within 1$\sigma$ uncertainty. This indicates that the contribution of the nebular continuum to the broad-band photometry of our sample is negligible and the massive nature of our sample is robust. 

To test the robustness of our results, we also use different SFH models (i.e., constant SFH, delayed SFH, and a combination of delayed SFH and a recent burst) and different codes (i.e., \texttt{BAGPIPES}\cite{Carnall2018} and \texttt{CIGALE}\cite{Boquien2019}) in the SED fitting for 36 DSFGs. The $M_{\star}$ and SFR values derived from different codes are generally consistent within the errors. However, there are global differences in SFR at different SFHs. Overall, the delayed SFH+burst model returns the highest SFR, followed by the constant SFH, and the delayed SFH returns the lowest SFR. This is reasonable because the SFR values here trace the most recent star formation (i.e., the last 10 Myr in \texttt{BAGPIPES}; instantaneous star formation in \texttt{CIGALE}). In addition, we note that the fitting results are strongly constrained by the range of initial parameter values, especially in the delayed SFH and the delayed SFH+burst models (i.e., a long timescale of decrease $\tau$ will yield a constant-like SFH and a short $\tau$ will lead to a burst-like SFH). Considering the optically dark nature of our sources, so that the photometric data points at UV-to-NIR wavelength are limited, we adopt the constant SFH -- the simplest and most constrained model -- in this work. In the Extended Data Table \ref{tab2}, we show the consistent $M_{\star}$ values derived from \texttt{BAGPIPES} and \texttt{CIGALE} for the three ultra-massive galaxies studied in this paper under different SFH models. 

Overall, with the JWST FRESCO imaging and grism spectroscopic survey, we present a robust stellar mass and redshift determination for a sample of DSFGs at $z_{\rm spec}=5-9$ (Fig.~\ref{fig2}). These galaxies are red (F182M $-$ F444W $>1.5$), with a wide distribution of log($M_{\rm \star}$/$M_{\odot}$) $=8.3-11.4$ and at $z_{\rm spec}>5$ (see Extended Data Table \ref{tab1}). Most intriguingly, our sample contains three spectroscopically confirmed ultra-massive galaxies (log$M_{\rm \star}$/$M_{\odot}$ $>11.0$) at $z_{\rm spec}\sim5.5$ and two ultra-massive galaxies (log$M_{\rm \star}$/$M_{\odot}$ $>10.4$) at $z_{\rm spec}\sim7.5$. They all present very efficient star formation with $\epsilon >0.2$ (Fig.~\ref{fig1}). To conclude, our findings of the existence of ultra-massive galaxies over a wide period (not only limited to the first few hundred million years), suggest that the Universe is forming galaxies much more efficiently than we expected.

In this study, we mainly focus on three ultra-massive galaxies at $z_{\rm spec}=5-6$ instead of two galaxies at $z_{\rm spec}>7$. The stellar masses of the latter are less robust, since the determination of stellar masses is mostly constrained by the F444W band. At $z\sim7-9$, this band traces the emission of the rest-frame B-band, which is sensitive to dust attenuation and stellar age. In addition, one of the two sources has a red point source morphology, indicating the existence of a potentially broad line AGN\cite{Matthee2024}. The other has only one emission line ($\sim$8$\sigma$), so the redshift and therefore the stellar mass may be uncertain. The error bars on the galaxies in Fig.~\ref{fig1} are just from SED fitting uncertainties and do not reflect these large systematic uncertainties for the two $z_{\rm spec}>7$ massive galaxies. At lower redshifts, the determination of the stellar mass and SFR is more robust because F444W traces the rest-frame optical.
\\
\\
\textbf{Reliability of the $z_{\rm spec}$ for S1.}
Among our three ultra-massive galaxies at $z_{\rm spec}=5-6$, S1 is the only source with a single emission line (14$\sigma$). Although the solution of the H$\alpha$ line with $z_{\rm H\alpha}=5.58$ is in excellent agreement with the $z_{\rm phot}=5.61_{-0.68}^{+1.44}$ from \texttt{EAZY}, we here also investigate the possibility of other emission lines. If the line with 14$\sigma$ detection was [OIII]$\lambda5008$ ($z_{\rm [OIII]}=7.62$) or [SIII]$\lambda9533$ ($z_{\rm [SIII]}=3.53$), we should also detect [OIII]$\lambda4960$ or [SIII]$\lambda9071$ over the wavelength coverage of F444W at about 5$\sigma$, considering the theoretical flux ratios of these two pairs ([OIII]$\lambda5008$/[OIII]$\lambda4960=2.92$ and [SIII]$\lambda9533$/[SIII]$\lambda9071=2.58$\cite{Landt2015}).  Over the wavelength coverage of the F444W filter (3.8 to 5.0 $\mu$m), the detection of only one line, i.e., the strongest line, can only be the Pa$\alpha$ ($z_{\rm Pa\alpha}=1.30$) line or Pa$\beta$ ($z_{\rm Pa\beta}=2.37$) line. Therefore, we mainly focus on the possibilities of H$\alpha$, Pa$\alpha$, and Pa$\beta$ in the following discussion. 

We perform SED fitting with \texttt{Bagpipes}, with the same parameter settings as mentioned above. In the fitting process, we fix the $z_{\rm spec}$ corresponding to the H$\alpha$, Pa$\alpha$, and Pa$\beta$ lines, respectively. The fitting results are shown in the Extended Data Fig.~\ref{solution}. The chi-square $\chi^2$ values for the H$\alpha$, Pa$\beta$, and Pa$\alpha$ lines are 1.9, 6.4, and 29.9, respectively. This shows that the best fit is obtained for the case of the H$\alpha$ line. 

We also test the different redshift solutions using \texttt{Bagpipes} with more degrees of freedom in the dust attenuation law. The CF00 dust attenuation law\cite{Charlot2000} combined with a wide Power-law slope value range from 0.5 to 2 returns $\chi^2$ values of 4.6, 12.3, and 47.5 for the H$\alpha$, Pa$\beta$, and Pa$\alpha$ lines, respectively. The Salim dust attenuation law\cite{Salim2018} combined with a slope value range from -1.0 to 0.5 returns $\chi^2$ values of 3.0, 6.0, and 22.6 for the H$\alpha$, Pa$\beta$, and Pa$\alpha$ lines, respectively.

In addition, since S1 has a strong detection at ALMA 1.1mm ($0.95\pm0.12$ mJy)\cite{Gómez-Guijarro2022}, we can calculate its SFR$_{\rm IR}$ under different redshift solutions, and then compare the SFR$_{\rm IR}$ with the SFR value derived from UV-to-NIR SED fitting (Extended Data Fig.~\ref{solution}). We calculate the SFR$_{\rm IR}$ based on the infrared luminosity, which is obtained from the IR template library\cite{Schreiber2018} normalized to the 1.1mm flux, following the method in the literature\cite{Xiao2023} (see section on Infrared luminosity and obscured star formation). For the H$\alpha$, Pa$\beta$, and Pa$\alpha$ line solutions, the derived SFR$_{\rm IR}$ values are $643\pm81\, M_{\odot}$yr$^{-1}$, $565\pm71\, M_{\odot}$yr$^{-1}$, and $377\pm48\, M_{\odot}$yr$^{-1}$, respectively, and the SFRs are $646^{+476}_{-199}\,M_{\odot}$yr$^{-1}$, $30^{+27}_{-8}\,M_{\odot}$yr$^{-1}$, and $1.9^{+0.4}_{-0.2}\,M_{\odot}$yr$^{-1}$, respectively. The SFR$_{\rm IR}$ and SFR values for the H$\alpha$ solution are in excellent agreement. However, for the Pa$\beta$ and Pa$\alpha$ solutions, their SFR$_{\rm IR}$ are about 20 and 200 times higher than their SFRs, respectively, resulting in a large discrepancy that cannot be explained by the uncertainties caused by the differences in the SED fitting models.

Furthermore, if we assume the observed line to be Pa$\beta$ (12821.6\AA), it is expected to be less affected by dust attenuation as it falls into the rest-frame near-infrared. Therefore, we can make a direct comparison between SFR$_\mathrm{Pa\beta}$ and SFR$_\mathrm{IR}$. We calculate the SFR$_\mathrm{Pa\beta}$ following the equation of SFR$_{\rm Pa\beta}$ [M$_{\odot}$yr$^{-1}$] = 3.84 $\times$ 10$^{-41}$ $\times$ $L$(Pa$\beta)$ [erg s$^{-1}$]\cite{Reddy2023}, where the $L$(Pa$\beta)$ is the luminosity of Pa$\beta$. The derived SFR$_{\rm Pa\beta}$ is $48\pm3\,M_{\odot}$yr$^{-1}$, which is about 12 times lower than the SFR$_{\rm IR}$.

Overall, all of the above tests provide poorer results to the Pa$\alpha$ and Pa$\beta$ lines compared to the H$\alpha$ line. Therefore, we rule out these two lines in S1. On the other hand, the solution of $z_{\rm spec}=5.58$ from the H$\alpha$ line is consistent with (1) the $z_{\rm phot}=5.61_{-0.68}^{+1.44}$ from \texttt{EAZY}; (2) the case of the excellent agreement between the SFR$_{\rm IR}$ and SFR values; (3) the evolutionary tracks of theoretical dusty star-forming galaxy templates, suggesting the presence of extremely dust-obscured massive galaxies at $z\gtrsim5$ (see Extended Data Fig.~\ref{Extended_color_mag}); and (4) the literature findings that optically dark/faint galaxies are generally located at $z>3$\cite{Wang2019, Zhou2020, Jin2022, Barrufet2023, Xiao2023}. 
Consequently, we consider the $z_{\rm spec}=5.58$ of H$\alpha$ line in S1 reliable.
\\
\\
\textbf{Possible AGN contamination.} To confirm the reliability of the ultra-massive nature of the three galaxies S1, S2, and S3, we further examined whether their line-subtracted broadband photometry could potentially be contaminated by AGN. 
The contamination of AGN emission could lead to an overestimation of stellar mass values in SED modeling that only accounts for stellar emission in models. Type II (narrow emission lines only) AGN can produce strong, narrow line emission, whose flux contribution has been securely subtracted from the broadband photometry before we perform the \texttt{Bagpipes} SED fitting, following the analysis procedure of typical star-forming galaxies. However, type I (broad emission line) AGNs also show continuum emission of the accretion disk, which cannot be removed with our method. Therefore, the most important thing is to check if there is a type I AGN in our three ultra-massive galaxies.

In our parent sample of 36 DSFGs, seven galaxies have distinct broad H$\alpha$ emission lines (full-width half maximum $v_{\rm FWHM, H\alpha, broad}>1,000$ km s$^{-1}$) with a red point source morphology, in agreement with the recently discovered ``little red dot (LRD)'' population\cite{Matthee2024, Kocevski2023, Labbe2023b}. These seven galaxies are presented in ref\cite{Matthee2024} and marked with ``LRD'' in the Extended Data Table \ref{tab1}. Among our three ultra-massive galaxies, none of them are in the sample of these seven galaxies, suggesting the lack of the typical telltale signs of type I AGN in their spectra (see Fig.~\ref{fig2}). Specifically, we note that S3 shows a relatively broader apparent line width compared to S1 and S2. However, the intrinsic $v_{\rm FWHM, H\alpha}$ of S3 is only found to be about 75 km s$^{-1}$. Its broad appearance in the 1D spectra is mainly due to the coincidence of alignment between the dispersion direction of the F444W grism and the elongated disk morphology of S3. Furthermore, if S3 has a strong type I AGN, we should see the H$\alpha$ line to be significantly wider than the [NII] line, but is not actually the case. On the other hand, we also investigate the structural properties of our three sources, to check whether they have a red point source morphology as type I AGN. The low values of best fit Sérsic index and extended morphology from \texttt{Galfit}\cite{Peng2002} suggest that they are not dominated by a point source (see the following section on Stellar structures of S1, S2, and S3 for detailed discussion). All of the evidence above suggests that the stellar masses of these galaxies are not significantly affected by an AGN and that their ultra-massive nature is reliable.
\\
\\
\textbf{Morphology of S1, S2, and S3}
We place constraints on the structural properties of our three sources in F444W and H$\alpha$ emission with \texttt{Galfit}\cite{Peng2002}. Images are fit with a single Sersic profile convolved with the PSF allowing the centroid, total brightness, half-light radius, Sersic index, axis ratio, and position angle to vary. PSFs are derived from the WebbPSF software\cite{Perrin2014} and rotated to match the position angle of the observations. Sources more than 3'' away or more than 2.5~mag fainter are masked; closer and brighter sources are fit simultaneously. All three sources have a best fit Sérsic index  $n<2.5$ and radii $R_{\rm e} > 3$ pixels, suggesting that they are not dominated by a point source (see Extended Data Table \ref{tab3} and Extended Data Fig.~\ref{galfit}). In order to quantify the potential contamination of their flux measurements by a Type I AGN, we compute the fraction of the total flux that remains in the central resolution element of the residual image. For all three sources, this is less than 2\%, suggesting that flux contamination from a potential AGN is not likely to be the driving factor behind the large fiducial mass estimates. 
\\
\\
\textbf{Infrared luminosity and obscured star formation.} S1, S2, and S3 are optically dark DSFGs that have star formation heavily obscured by dust, with $A_{\rm V}>3$ mag. Although we have obtained their SFRs from UV-to-NIR SED fits corrected for dust attenuation (Extended Data Table \ref{tab1}), we independently calculated their obscured star formation here using infrared wavelength data (SFR$_{\rm IR}$). Considering that the three sources have different far-infrared datasets, we used different methods to obtain their total infrared luminosity ($L_{\rm IR}$; 8-1000$\mu$m rest-frame) and SFR$_{\rm IR}$.

For S1, we only have the 850$\mu$m flux from the JCMT/SCUBA-2 and ALMA band-7 observations\cite{Cowie2018}, and the 1.1mm flux from the GOODS-ALMA survey\cite{Gómez-Guijarro2022, Xiao2023}. For S3, we have 450$\mu$m and 850$\mu$m fluxes from JCMT/SCUBA-2\cite{Cowie2017, Barger2022}. These two sources are either not detected in Herschel or are strongly affected by neighboring bright sources, so we do not include Herschel values in our analysis here.

We first perform the FIR SED fitting to S3 with \texttt{CIGALE} using fixed $z_{\rm spec}$. We use the Draine dust emission templates\cite{Draine2014} to derive $L_{\rm IR}$, with the same parameter settings as in ref.\cite{Xiao2023}. The SFR$_{\rm IR}$ is then calculated based on the $L_{\rm IR}$, following the method of ref.\cite{Kennicutt2012}. The derived values for S3 are $L_{\rm IR}=(6.6\pm0.3) \times 10^{12}\, L_{\odot}$ and SFR$_{\rm IR}=988\pm49\, M_{\odot}$yr$^{-1}$. In addition, for S1, due to the lack of photometric constraints around the peak of FIR SED ($\sim$500$\mu$m at the observed frame given the redshift of S1), we use two different approaches for the SED fitting analysis. The first method is to perform the SED fitting using \texttt{CIGALE} with fixed $z_{\rm spec}$ as described above. The second method is to re-normalize the IR templates\cite{Schreiber2018} to the ALMA 1.1mm flux, according to the method of ref.\cite{Xiao2023}. The two methods yielded consistent values, i.e., $L_{\rm IR}=(5.3\pm0.3) \times 10^{12}\, L_{\odot}$ and SFR$_{\rm IR}=795\pm40\, M_{\odot}$yr$^{-1}$ from the first methods and $L_{\rm IR}=(4.3\pm0.6) \times 10^{12}\, L_{\odot}$ and SFR$_{\rm IR}=643\pm81\, M_{\odot}$yr$^{-1}$ from the second methods. In the main body of the paper, we show the values obtained by the first method.

For S2, the so-called GN10, we obtain its $L_{\rm IR}=1.03_{-0.15}^{+0.19} \times 10^{13}\, L_{\odot}$ and SFR$_{\rm IR}=1,030_{-150}^{+190}\, M_{\odot}$yr$^{-1}$ from ref.\cite{Riechers2020}. These values are also derived from \texttt{CIGALE} SED fits, based on extensive data from Herschel, JCMT/SCUBA, SMA, and NOEMA observations.

The SFR$_{\rm IR}$ of S1 is similar to the SFR value derived from the UV-to-NIR SED fit (Extended Data Table \ref{tab1}). However, the SFR$_{\rm IR}$ of S2 and S3 are about three times higher than their SFR$_{\rm SED}$, implying that there may be hidden dust regions in these galaxies that absorb all the UV photons, which cannot be reproduced with a dust extinction correction\cite{Xiao2023, Elbaz2018, Puglisi2017}, or that our assumed star-formation histories are not appropriate. Irrespective of this, the high SFR$_{\rm IR}$ values indicate that our galaxies are in a very efficient mass assembly and accumulation process, which explains why they are so massive.
\\
\\
\textbf{The maximal halo mass ($M_\mathrm{halo}^\mathrm{max}$).} In this paper, we define the $M_\mathrm{halo}^\mathrm{max}$ as the mass above which the cumulative dark matter halo mass function predicts one halo to be detected in the FRESCO survey volume ($V$). Under the most updated Planck cosmology\cite{Planck2020}, we consider the full FRESCO survey volume as the comoving cube enclosed by the 124 arcmin$^2$ survey area between $z=4.9$ and $z=9.0$, where either H$\alpha$+[NII] or H$\beta$+[OIII] fall into the spectral coverage. This corresponds to a total survey volume of $\sim1.2\times10^{6}$ comoving Mpc$^3$. As for the halo mass function, we compute it following \cite{Tinker2008} using the Python package \texttt{hmf}\cite{Murray2013} under the same Planck cosmology\cite{Planck2020}. With the halo mass function, we further derive the cumulative number density of dark matter halos above a certain mass ($n$), as a function of redshift and halo mass. According to the definition, the maximum halo mass at a given redshift (z) thus satisfies n(M$_\mathrm{halo}^\mathrm{max}$, $z$) $\times$ $V$ = 1, from which we derive its redshift evolution and further constrain the maximal stellar mass of galaxies expected by very simplistic galaxy assembly models in Fig.~\ref{fig1}.
\\
\\
\textbf{Cosmic variance.} Considering the limited and non-continuous sky area covered by the FRESCO survey (62 arcmin$^2$ each for the GOODS-North and -South fields), we include the effect of cosmic variance in this work. The cosmic variance is derived from the FLAMINGO simulations\cite{Schaye2023, Kugel2023}. We adopt the Dark Matter Only (DMO) simulation with a box size of 2.8 Gpc. It is the largest box available while having a high-resolution dark matter particle mass ($6.72\times10^{9} M_{\odot}$), making it the best choice for our study. In total, we have 12 redshift snapshots between $z\sim4-10$ at intervals of $\sim0.5$. In each redshift, we obtain the most massive halo mass from each of the 32,768 subboxes of FLAMINGO. Every subbox has a volume matching half of the total FRESCO survey volume ($\sim1.2\times10^{6}$ cMpc$^3$), considering the real observed GOODS-North and -South fields are not contiguous. Then, we perform a bootstrap approach to select the most massive halo mass among any two random subboxes. This procedure is repeated 10,000 times, and we calculate the 16th and 84th percentiles of the distribution of the maximum halo mass as 1$\sigma$ uncertainty of cosmic variance.

\begin{addendum}

 \item[Data availability] 
 All the raw data are publicly available through the \texttt{Mikulski Archive for Space Telescopes}\footnote{\url{https://archive.stsci.edu/}} (\texttt{MAST}), under program ID 1895.
The FRESCO data are being released on MAST as a High-Level Science Product via \url{https://doi.org/10.17909/gdyc-7g80}. Images are already available. The spectra are being calibrated and will be discussed in an upcoming data paper (Brammer et al., in prep.). For updates, please check the survey webpage: \url{https://jwst-fresco.astro.unige.ch/} or the MAST page \url{https://archive.stsci.edu/hlsp/fresco/}. The advanced datasets generated during and/or analysed during the current study are available from the corresponding author on reasonable request.

 \item[Code availability] 
 The codes used to reduce and analyze data in this work are publicly available: \texttt{grizli} (\url{https://github.com/gbrammer/grizli}), \texttt{EAZY} (\url{https://github.com/gbrammer/eazy-photoz}),  \texttt{Bagpipes} (\url{https://github.com/ACCarnall/bagpipes}), \texttt{CIGALE} (\url{https://cigale.lam.fr/}), IR template library (\url{http://cschreib.github.io/s17-irlib/}), \texttt{Galfit} (\url{https://users.obs.carnegiescience.edu/peng/work/galfit/galfit.html}).
 
 \item[Acknowledgements] The authors thank R. Marques-Chaves for helpful discussions. This work is based on observations made with the NASA/ESA/CSA James Webb Space Telescope. The data were obtained from the Mikulski Archive for Space Telescopes at the Space Telescope Science Institute, which is operated by the Association of Universities for Research in Astronomy, Inc., under NASA contract NAS 5-03127 for JWST. These observations are associated with program \# 1895. Support for this work was provided by NASA through grant JWST-GO-01895 awarded by the Space Telescope Science Institute, which is operated by the Association of Universities for Research in Astronomy, Inc., under NASA contract NAS 5-26555. This work has received funding from the Swiss State Secretariat for Education, Research and Innovation (SERI) under contract number MB22.00072, as well as from the Swiss National Science Foundation (SNSF) through project grant 200020\_207349.
The Cosmic Dawn Center (DAWN) is funded by the Danish National Research Foundation under grant DNRF140. RPN acknowledges funding from JWST programs GO-1933 and GO-2279. Support for this work was provided by NASA through the NASA Hubble Fellowship grant HST-HF2-51515.001-A awarded by the Space Telescope Science Institute, which is operated by the Association of Universities for Research in Astronomy, Incorporated, under NASA contract NAS5-26555. YF acknowledges support from NAOJ ALMA Scientific Research Grant number 2020-16B. YQ acknowledges support from the Australian Research Council Centre of Excellence for All Sky Astrophysics in 3 Dimensions (ASTRO 3D), through project number CE170100013. IL acknowledges support by the Australian Research Council through Future Fellowship FT220100798. MS acknowledges support from the CIDEGENT/2021/059 grant, from project PID2019-109592GB-I00/AEI/10.13039/501100011033 from the Spanish Ministerio de Ciencia e Innovaci\'on - Agencia Estatal de Investigaci\'on. MST also acknowledges the financial support from the MCIN with funding from the European Union NextGenerationEU and Generalitat Valenciana in the call Programa de Planes Complementarios de I+D+i (PRTR 2022) Project (VAL-JPAS), reference ASFAE/2022/025. Cloud-based data processing and file storage for this work is provided by the AWS Cloud Credits for Research program. This work used the DiRAC@Durham facility managed by the Institute for Computational Cosmology on behalf of the STFC DiRAC HPC Facility (www.dirac.ac.uk). The equipment was funded by BEIS capital funding via STFC capital grants ST/K00042X/1, ST/P002293/1, ST/R002371/1, and ST/S002502/1, Durham University and STFC operations grant ST/R000832/1. DiRAC is part of the National e-Infrastructure.

  \item[Author Contributions] 
 M.X. performed the majority of the present analysis and wrote the majority of the text, with significant help from P.A.O., D.E., and L.B. E.N. performed Galfit modeling of the sources. A.W. produced the multi-wavelength catalogs used in this work. G.D.I. and P.D. support the scientific interpretation.  G.B. reduced the JWST/NIRCam images and grism data. S.L. helped with the FLAMINGO simulation. All authors contributed to the manuscript and helped with the analysis and interpretation.
 
 \item[Author Information] The authors declare that they have no competing financial interests. Correspondence and requests for materials should be addressed to M.X. (mengyuan.xiao@unige.ch).

\end{addendum}

\let\oldthebibliography=\thebibliography
\let\oldendthebibliography=\endthebibliography
\renewenvironment{thebibliography}[1]{%
    \oldthebibliography{#1}%
    \setcounter{enumiv}{42} %change this number to the number of last citation in the main text
}{\oldendthebibliography}
\putbib[only_in_appendix]

\onecolumn

\renewcommand{\arraystretch}{0.6}
\renewcommand{\baselinestretch}{1}

\newpage
\onecolumn
\noindent\textbf{\large Extended Data}
\renewcommand\thefigure{\arabic{figure}}
\setcounter{figure}{0}
\renewcommand{\tablename}{Extended Data Table}
%\begin{table}[h!]
\begin{table*}[!tbh]\centering\begin{minipage}{1.\textwidth}\footnotesize %\scriptsize
  \centering
  \caption{\small{\textbf{Physical properties of the whole sample of 36 DSFGs in FRESCO-S\&N fields at $z_{\rm spec} \gtrsim5$.} (1) IDs; (2) Source ID in FRESCO; (3)(4) Right ascension and declination (J2000) of sources from JWST; (5) Spectroscopic redshift identified from H${\alpha}$ ($z_\mathrm{{spec}}=$ 4.86 - 6.69) or [OIII] ($z_\mathrm{{spec}}=$ 6.69 - 9.08) emission lines. For sources with only one emission line detected at $>8\sigma$ (flagged with a ``*'' exponent), their $z_{\rm spec}$ must be within the 16-84th percentile uncertainties of the $z_{\rm phot}$;  (6) signal-to-noise ratio (S/N) of H${\alpha}$ or [OIII] line; (7)(8)(9) $M_{\rm \star}$, SFR, and $A_{\rm V}$ derived from UV-to-NIR SED fitting at a fixed $z_{\rm spec}$ assuming a constant SFH and a Chabrier IMF\cite{Chabrier2003}; (10) AGN candidates, showing broad H$\alpha$ emission lines ($v_{\rm FWHM, H\alpha, broad}>1000$ km s$^{-1}$) with a ``little red dot (LRD)'' morphology\cite{Matthee2024}; (11) Source IDs in other work. All uncertainties refer to 1$\sigma$ standard deviation.
}}
  \medskip
  \begin{tabular}{l l c clc rr r c c }
    \hline
    \hline
    \noalign{\medskip}
S.No. & ID & RA & DEC & $z_\mathrm{{spec}}$ & S/N$_{\rm H\alpha\, or\, [OIII]}$ & log($M_{\rm \star}$) & log(\rm{SFR})  & $A_{\rm V}$ & AGN & other ID \\ 
& & (deg) & (deg) & & &log($M_{\odot}$) &  log($M_{\odot}$yr$^{-1}$)   & (mag)\\ 
(1)& (2) & (3) & (4)&(5)&(6)&(7)&(8)&(9)&(10)& (11)\\
	\noalign{\medskip}
	\hline
	\noalign{\smallskip}

   & S-15496 & 03:32:24.66 & -27:44:40.08  &  5.049$^{*}$ &17.2&   9.72$_{-0.12}^{+0.12}$ &   1.05$_{-0.16}^{+0.22}$  &  1.7$_{-0.2}^{+0.3}$ & \\
   & S-11579 & 03:32:20.82 & -27:47:14.76  &  5.330$^{*}$ &20.9&   9.44$_{-0.15}^{+0.11}$ &   0.82$_{-0.14}^{+0.19}$  &  1.4$_{-0.2}^{+0.3}$ & \\
   & S-4166  & 03:32:35.44 & -27:50:31.38  &  5.402$^{*}$ &19.6&   9.14$_{-0.09}^{+0.08}$ &   0.42$_{-0.08}^{+0.11}$  &  0.2$_{-0.1}^{+0.1}$ & \\
   & S-10944 & 03:32:27.46 & -27:47:31.63  &  5.445 &9.14&   8.64$_{-0.15}^{+0.15}$ &   0.41$_{-0.15}^{+0.14}$  &  0.3$_{-0.2}^{+0.2}$ & &JADES-116930\cite{Bunker2023} \\
   & S-11150 & 03:32:33.26 & -27:47:24.91  &  5.483 &30.4&   9.96$_{-0.06}^{+0.05}$ &   1.17$_{-0.05}^{+0.06}$  &  1.0$_{-0.1}^{+0.1}$ &&JADES-204851\cite{Bunker2023} \\
S1 & S-18258 & 03:32:28.91 & -27:44:31.53  &  5.579$^{*}$ &14.3&  11.37$_{-0.13}^{+0.11}$ &   2.81$_{-0.16}^{+0.24}$  &  3.2$_{-0.2}^{+0.3}$  & & OFG28\cite{Xiao2023}, A2GS33\cite{Gómez-Guijarro2022}, ID68\cite{Cowie2018}, ID20\cite{Yamaguchi2019} \\
   & S-5661  & 03:32:20.87 & -27:49:54.85  &  5.774$^{*}$ &10.6&   9.56$_{-0.12}^{+0.12}$ &   1.00$_{-0.14}^{+0.22}$  &  1.4$_{-0.2}^{+0.3}$ & \\
   & S-12686 & 03:32:34.63 & -27:46:47.48  &  6.090$^{*}$ &15.5&   8.29$_{-0.13}^{+0.15}$ &   0.33$_{-0.14}^{+0.11}$  &  1.0$_{-0.2}^{+0.2}$ & \\
   & S-19392 & 03:32:38.73 & -27:44:15.60  &  6.815 &20.9&   9.97$_{-0.07}^{+0.06}$ &   1.30$_{-0.07}^{+0.08}$  &  1.1$_{-0.1}^{+0.1}$ && JADES-219000\cite{Bunker2023} \\
   & S-6209  & 03:32:32.12 & -27:49:41.72  &  7.664 &8.3 &  10.48$_{-0.17}^{+0.18}$ &   2.05$_{-0.20}^{+0.22}$  &  3.7$_{-0.4}^{+0.4}$ & & JADES-90354\cite{Bunker2023} \\
   & S-8010  & 03:32:20.99 & -27:48:53.71  &  7.846 &26.8&   9.30$_{-0.14}^{+0.12}$ &   0.88$_{-0.12}^{+0.17}$  &  0.3$_{-0.1}^{+0.2}$ & &JADES-GS-53.08745-27.81492\cite{Hainline2024} \\
\hline 
   & N-1002   &  12:36:35.50 & +62:11:04.29  &  5.050$^{*}$ &10.9&   9.37$_{-0.17}^{+0.13}$ & 0.82$_{-0.17}^{+0.27}$ &    1.4$_{-0.3}^{+0.4}$ &  \\
   & N-15498  &  12:37:08.53 & +62:16:50.82  &  5.086 &17.0&  10.12$_{-0.13}^{+0.11}$ & 1.37$_{-0.14}^{+0.15}$ &    2.0$_{-0.2}^{+0.2}$ &  LRD \\
   & N-6079   &  12:36:55.73 & +62:13:36.33  &  5.093$^{*}$ &15.1&   8.98$_{-0.11}^{+0.08}$ & 0.27$_{-0.10}^{+0.13}$ &    0.6$_{-0.1}^{+0.1}$ &   \\
   & N-14409  &  12:36:17.30 & +62:16:24.35  &  5.146 &30.6&   9.09$_{-0.09}^{+0.09}$ & 0.33$_{-0.09}^{+0.08}$ &    0.2$_{-0.1}^{+0.1}$ &  LRD\\
   & N-1257   &  12:36:24.68 & +62:11:17.01  &  5.167 &21.8&  10.04$_{-0.19}^{+0.14}$ & 1.82$_{-0.16}^{+0.14}$ &    1.6$_{-0.2}^{+0.2}$ &   \\
S3 & N-2663   &  12:36:56.56 & +62:12:07.37  &  5.179 &22.6&  11.04$_{-0.18}^{+0.17}$ & 2.48$_{-0.20}^{+0.29}$ &    3.4$_{-0.4}^{+0.6}$ &   \\
   & N-3188   &  12:36:51.97 & +62:12:26.04  &  5.187 &15.7&  10.58$_{-0.17}^{+0.14}$ & 2.30$_{-0.17}^{+0.18}$ &    1.8$_{-0.2}^{+0.2}$ & & HDF850.1\cite{Walter2012, Herard-Demanche2023} \\
   & N-7162   &  12:37:16.90 & +62:14:00.90  &  5.189 &21.7&  10.32$_{-0.14}^{+0.11}$ & 1.71$_{-0.12}^{+0.15}$ &    1.6$_{-0.1}^{+0.2}$ &  \\
   & N-16116  &  12:36:56.62 & +62:17:07.97  &  5.194$^{*}$ &14.6&   9.27$_{-0.08}^{+0.09}$ & 0.51$_{-0.08}^{+0.09}$ &    0.5$_{-0.1}^{+0.1}$ &  \\
   & N-4014   &  12:37:12.03 & +62:12:43.36  &  5.221 &34.1&   9.76$_{-0.10}^{+0.09}$ & 1.05$_{-0.11}^{+0.09}$ &    1.2$_{-0.1}^{+0.1}$ & LRD \\           
   & N-12839  &  12:37:22.63 & +62:15:48.11  &  5.240 &40.2&   9.94$_{-0.09}^{+0.08}$ & 1.16$_{-0.08}^{+0.08}$ &    1.0$_{-0.1}^{+0.1}$ & LRD \\
   & N-13733  &  12:36:13.70 & +62:16:08.18  &  5.243 &25.0&   9.54$_{-0.11}^{+0.09}$ & 0.77$_{-0.11}^{+0.12}$ &    1.0$_{-0.2}^{+0.2}$ & LRD \\
S2 & N-7496   &  12:36:33.42 & +62:14:08.57  &  5.306 &14.3&  11.18$_{-0.17}^{+0.23}$ & 2.58$_{-0.18}^{+0.30}$ &    3.4$_{-0.5}^{+0.8}$ &  & GN10\cite{Riechers2020}\\
   & N-16813  &  12:36:43.03 & +62:17:33.12  &  5.359 &48.5&   9.38$_{-0.11}^{+0.07}$ & 0.67$_{-0.03}^{+0.05}$ &    0.1$_{-0.0}^{+0.1}$ &   LRD\\
   & N-9771   &  12:37:07.44 & +62:14:50.31  &  5.535 &101.9& 10.24$_{-0.19}^{+0.20}$ & 2.05$_{-0.15}^{+0.12}$ &    1.8$_{-0.2}^{+0.2}$ & LRD \\
   & N-6924   &  12:37:02.72 & +62:13:55.07  &  5.535$^{*}$ &20.3&   9.38$_{-0.08}^{+0.07}$ & 0.64$_{-0.08}^{+0.08}$ &    0.3$_{-0.1}^{+0.1}$ &  \\
   & N-14178  &  12:36:41.69 & +62:16:18.39  &  5.569$^{*}$ &14.7&   8.90$_{-0.25}^{+0.30}$ & 0.35$_{-0.29}^{+0.37}$ &    1.2$_{-0.5}^{+0.9}$ &  \\
   & N-1459   &  12:36:41.95 & +62:11:24.47  &  6.024$^{*}$ &15.7&   9.46$_{-0.09}^{+0.07}$ & 0.78$_{-0.08}^{+0.09}$ &    0.4$_{-0.1}^{+0.1}$ &  \\
   & N-12207  &  12:36:36.47 & +62:15:34.72  &  6.760$^{*}$ &23.5&   9.75$_{-0.11}^{+0.09}$ & 1.11$_{-0.09}^{+0.11}$ &    0.6$_{-0.1}^{+0.1}$ &  \\
   & N-9094   &  12:36:04.62 & +62:14:36.71  &  7.038 &73.7&  10.48$_{-0.12}^{+0.11}$ & 1.91$_{-0.12}^{+0.15}$ &    1.3$_{-0.1}^{+0.2}$ &  \\
   & N-489    &  12:36:47.52 & +62:10:37.27  &  7.131 &31.4&   9.16$_{-0.19}^{+0.15}$ & 0.80$_{-0.20}^{+0.24}$ &    0.7$_{-0.3}^{+0.3}$ &  \\
   & N-8697   &  12:37:00.04 & +62:14:29.00  &  7.146 &19.3&   9.17$_{-0.13}^{+0.12}$ & 0.69$_{-0.15}^{+0.17}$ &    0.4$_{-0.2}^{+0.2}$ &  \\
   & N-2756   &  12:36:20.04 & +62:12:09.29  &  7.190 &18.4&  10.00$_{-0.10}^{+0.08}$ & 1.42$_{-0.11}^{+0.11}$ &    0.8$_{-0.1}^{+0.1}$ &  \\
   & N-11316  &  12:36:10.84 & +62:15:16.32  &  7.403$^{*}$ &14.2&   9.55$_{-0.14}^{+0.11}$ & 1.09$_{-0.16}^{+0.20}$ &    0.8$_{-0.2}^{+0.3}$ &  \\
   & N-4380   &  12:36:41.09 & +62:12:53.12  &  8.613 &16.4&   9.54$_{-0.15}^{+0.16}$ & 1.14$_{-0.16}^{+0.21}$ &    1.0$_{-0.2}^{+0.3}$ &  & JADES-GN-189.17121+62.21476\cite{Hainline2024}\\
\hline 

\hline
    \end{tabular}
    \label{tab1}
%\end{table}
\end{minipage}
\end{table*} 

\renewcommand{\arraystretch}{1.0}
\renewcommand{\baselinestretch}{1.5}

%\begin{table}[h!]
\begin{table*}[!tbh]
      \caption{\small{\textbf{Stellar masses of S1, S2, and S3.} We use the \texttt{Bagpipes} and \texttt{CIGALE} SED fitting codes, assuming a constant SFH (CSFH), a delayed SFH (DSFH), and a combination of delayed SFH and a recent burst  (DSFH+burst), respectively. All uncertainties refer to 1$\sigma$ standard deviation.
}}
    \centering 
        \begin{tabular}{crrrrrrrr}%\toprule
            \hline
            \hline
            & \multicolumn{3}{c}{\textbf{\texttt{Bagpipes}}} &\multicolumn{3}{c}{\textbf{\texttt{CIGALE}}} 
            
            \\\ S.No.  & CSFH& DSFH & DSFH+burst & CSFH  & DSFH & DSFH+burst \\
            \hline
            S1 & 11.37$_{-0.13}^{+0.11}$ & 11.34$_{-0.13}^{+0.11}$ & 11.33$_{-0.12}^{+0.11}$ & 11.16$_{-0.22}^{+0.15}$ & 11.27$_{-0.16}^{+0.11}$ & 11.42$_{-0.32}^{+0.18}$ \\
            S2 & 11.18$_{-0.17}^{+0.23}$ & 11.17$_{-0.21}^{+0.21}$ & 11.14$_{-0.20}^{+0.22}$ & 11.11$_{-0.32}^{+0.18}$ & 11.40$_{-0.10}^{+0.08}$ & 11.36$_{-0.07}^{+0.06}$  \\
            S3 & 11.04$_{-0.18}^{+0.17}$ & 11.01$_{-0.15}^{+0.16}$ & 10.96$_{-0.18}^{+0.18}$ & 10.99$_{-0.06}^{+0.05}$ & 11.00$_{-0.12}^{+0.09}$ & 11.04$_{-0.16}^{+0.12}$ \\
            \hline
            %\bottomrule
        \end{tabular}
        \label{tab2}
    \end{table*}

\begin{table*}[!tbh]
      \caption{\small{\textbf{Structural parameters of S1, S2, and S3.} Half light radii ($R_{\rm e}$), sérsic indices (n), and minor over major axis ratios (q) in the F444W direct image and the grism image. Though much of the structure observed in the grism image is due to the H$\alpha$ morphology, some may also be contributed by the kinematic structure and the [NII] line. The spatially extended structures and similarity of the morphologies between the grism and direct images for each object suggest that the light of these objects is not dominated by AGN. All uncertainties refer to 1$\sigma$ standard deviation.
}}
    \centering 
        \begin{tabular}{ccccccc}%\toprule
            \hline
            \hline
            & \multicolumn{3}{c}{\textbf{{direct image}}} &\multicolumn{3}{c}{\textbf{{H$\alpha$}}} 
            
            \\\ S.No.  & $R_{\rm e}$ (kpc) & n & q & $R_{\rm e}$ (kpc)  & n & q \\
            \hline
        S1 & 0.76$\pm$0.01 & 1.57$\pm$0.03 & 0.54$\pm$0.01 & 0.89$\pm$0.08 & 0.20$\pm$0.19 & 0.72$\pm$0.08 \\
        S2 & 2.57$\pm$0.25 & 1.18$\pm$0.14 & 0.64$\pm$0.04 & 1.92$\pm$0.46 & 1.55$\pm$0.43 & 0.77$\pm$0.13 \\
        S3 & 2.22$\pm$0.09 & 2.31$\pm$0.09 & 0.42$\pm$0.01 & 2.10$\pm$0.09 & 0.20$\pm$0.11 & 0.18$\pm$0.02 \\
            \hline
            %\bottomrule
        \end{tabular}
        \label{tab3}
    \end{table*}

\clearpage

\begin{Extended Data Figure}
	\centering
	\includegraphics[width=1.0\textwidth]{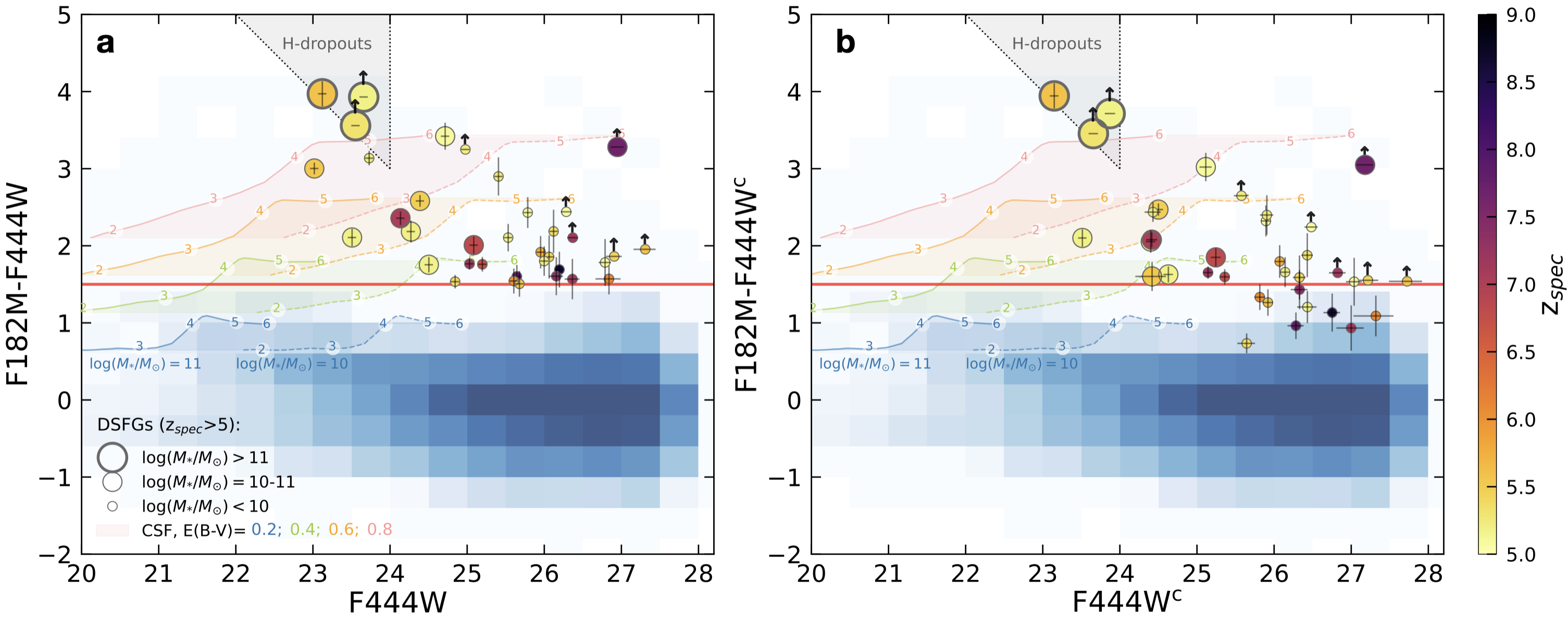} 
 	\caption{\small{\textbf{Color-magnitude diagram of 36 DSFGs color-coded by $z_{\rm spec}$.} 
F444W$^{\rm c}$ (\textbf{b}) is the corrected magnitude of F444W (\textbf{a}) at 4.44$\mu$m after subtracting the emission line contribution.  
Our sample of DSFGs (circles) at $z_{\rm spec} \gtrsim5$ is selected with F182M $-$ F444W $>$ 1.5 mag (red line). 
Galaxies with different $M_{\star}$ show different sizes in circles. 
All the DSFGs have $>$5$\sigma$ detections at F444W. 
For those undetected at F182M, we present their 2$\sigma$ limits with arrows. Error bars correspond to 1$\sigma$ uncertainties. 
The blue-shaded region describes the distribution of all F444W-detected sources in the FRESCO survey. For comparison, the grey triangular region shows traditional selection criteria of $H$-dropouts\cite{Wang2019} before the JWST spectral era. We here assume that the HST $H$-band and \textit{Spitzer}/IRAC 4.5\,$\mu$m band used in the pre-JWST selection criteria are approximate to those of JWST F182M and F444W, respectively.  We overlap the evolutionary tracks of theoretical dusty star-forming galaxy templates (blue, green, orange, and red regions tracing different degrees of reddening) at $z=2-6$. The numbers on the tracks indicate the redshift. The solid and dashed tracks correspond to log($M_{\star}/M_{\odot}$) = 11 and 10, respectively.
}}
	\label{Extended_color_mag}
\end{Extended Data Figure}

\clearpage

\begin{Extended Data Figure*}
	\centering
	\includegraphics[width=0.7\textwidth]{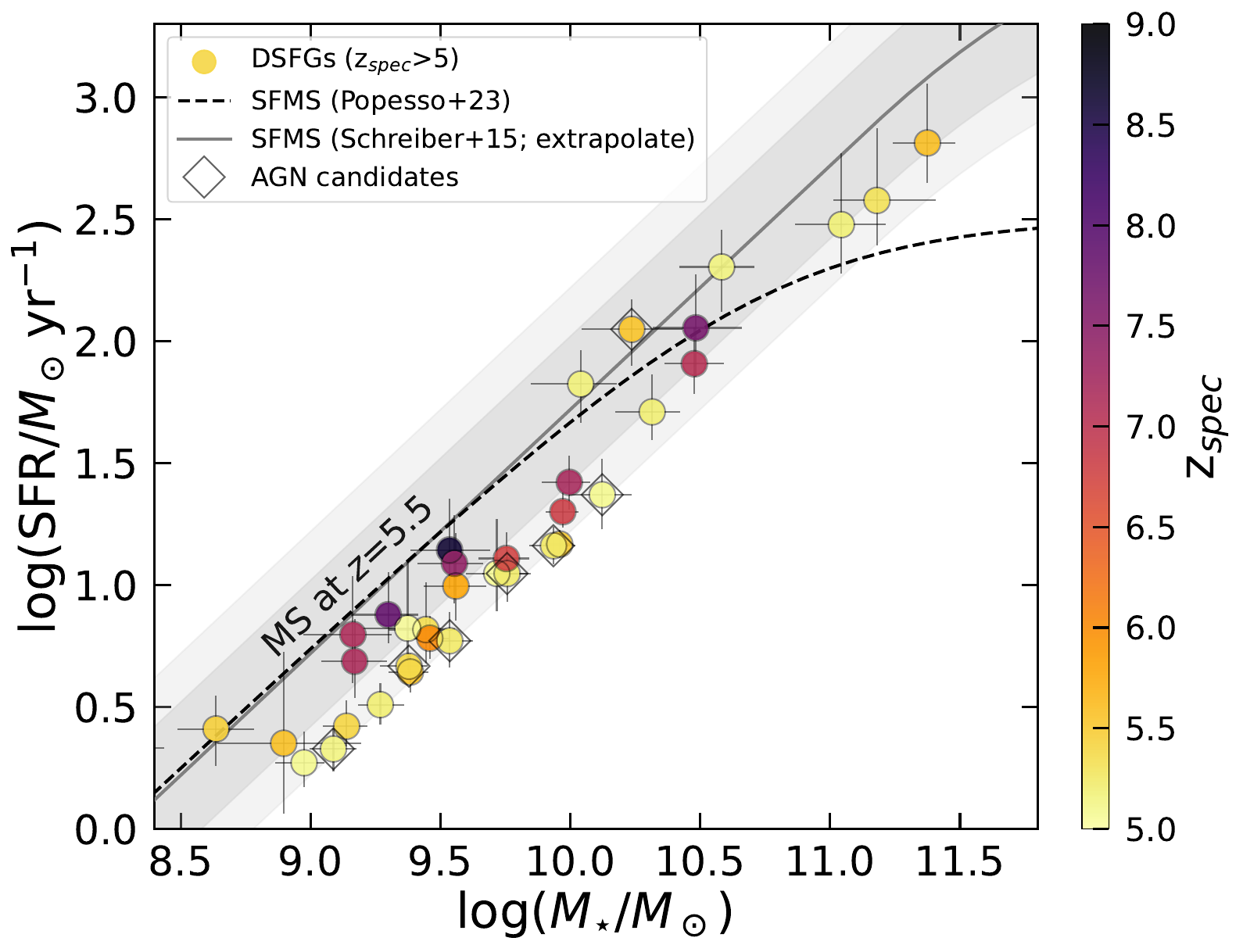} 
	\caption{\small{\textbf{Locations of 36 DSFGs compared to the SFMS in the SFR-$M_{\star}$ plane.}  Error bars correspond to 1$\sigma$ uncertainties.  The extrapolated Schreiber SFMS\cite{Schreiber2015} at $z=5.5$, 1$\sigma$ scatter (0.5 $<$ $\Delta$MS $<$ 2, i.e., $\sim$0.3 dex), and $\pm 3 \times \Delta$MS region (0.33 $<$ $\Delta$MS $<$ 3, i.e., $\sim0.5$ dex) are highlighted with a grey line, a dark grey shaded area, and a light grey shaded area, respectively. $\Delta$MS $>$ 3 is commonly used to separate MS and SB galaxies. We also plot the Popesso SFMS\cite{Popesso2023} at $z=5.5$  with a dashed line.
}}
	\label{Extended_MS}
\end{Extended Data Figure*}

\begin{Extended Data Figure}
	\centering
	\includegraphics[width=1.01\textwidth]{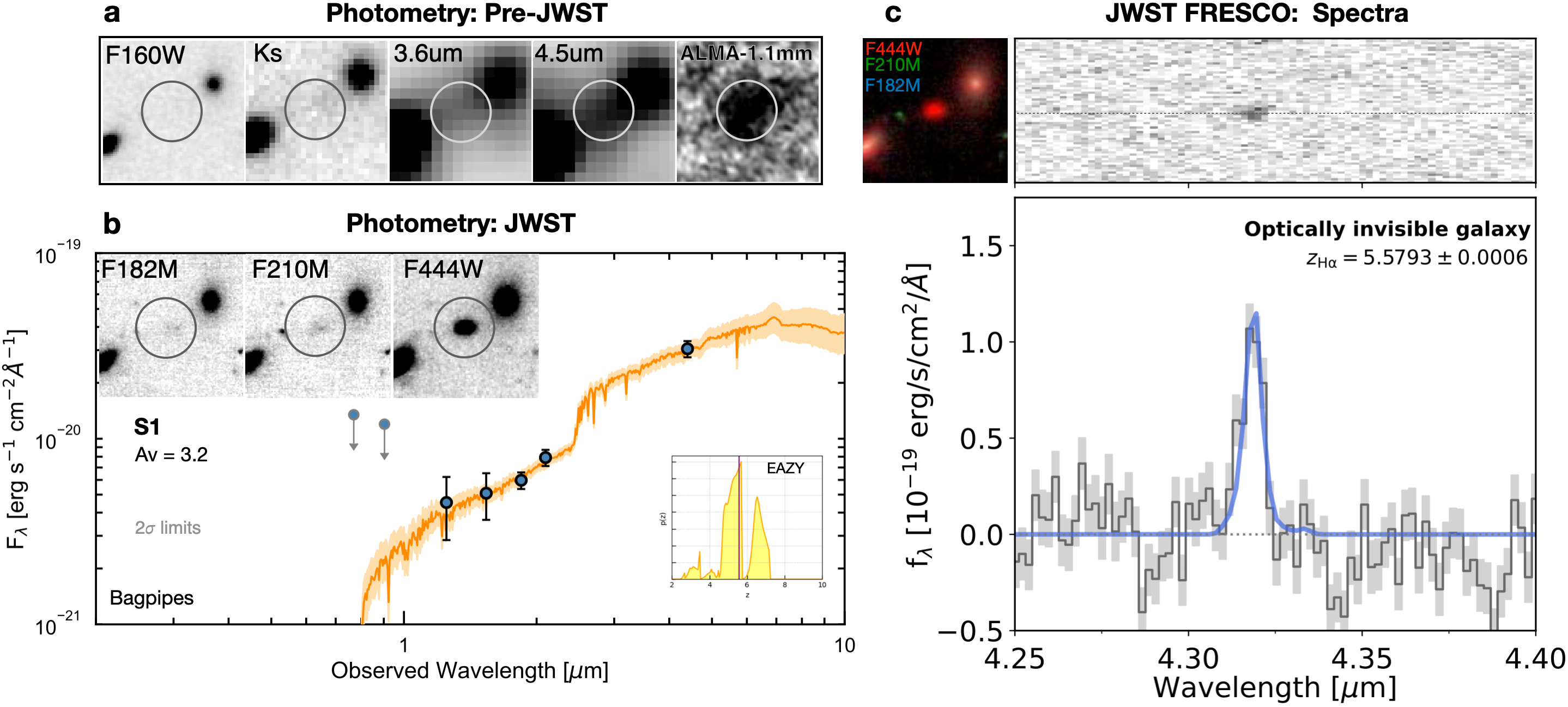} 
	\caption{\small{\textbf{The most extreme source S1: an optically dark, ultra-massive galaxy at $z_{\rm spec} = 5.58$.}  \textbf{a}: 4$^{\prime\prime}$ $\times$ 4$^{\prime\prime}$ stamp images of HST/WFC3 (F160W)\cite{Whitaker2019}, ZFOURGE ($K_{\rm s}$)\cite{Straatman2016}, \textit{Spitzer}/IRAC (3.6 $\mu$m \& 4.5 $\mu$m)\cite{Stefanon2021}, and ALMA band 6 (1.1 mm)\cite{Gómez-Guijarro2022, Xiao2023}. Before the JWST era, this source was only detected by the (sub)millimeter observations\cite{Cowie2018, Yamaguchi2019, Gómez-Guijarro2022, Xiao2023}. \textbf{b}: Stamps from JWST FRESCO survey with three filters. We also show SED fitting results from \texttt{Bagpipes} with fixed $z_{\rm spec}$ and the likelihood distributions of $z_{\rm phot}$ from EAZY, which perfectly match the $z_{\rm spec}$. Error bars correspond to 1$\sigma$ uncertainties, and downward arrows represent 2$\sigma$ upper limits. \textbf{c}: 2D (top)  and 1D (bottom) spectra from the FRESCO NIRCam/grism (F444W filter), with the Gaussian fit of the H${\alpha}$ emission line and the other lines around it.
    }}
	\label{Extended_S1}
\end{Extended Data Figure}

\begin{Extended Data Figure}
	\centering
	\includegraphics[width=1.01\textwidth]{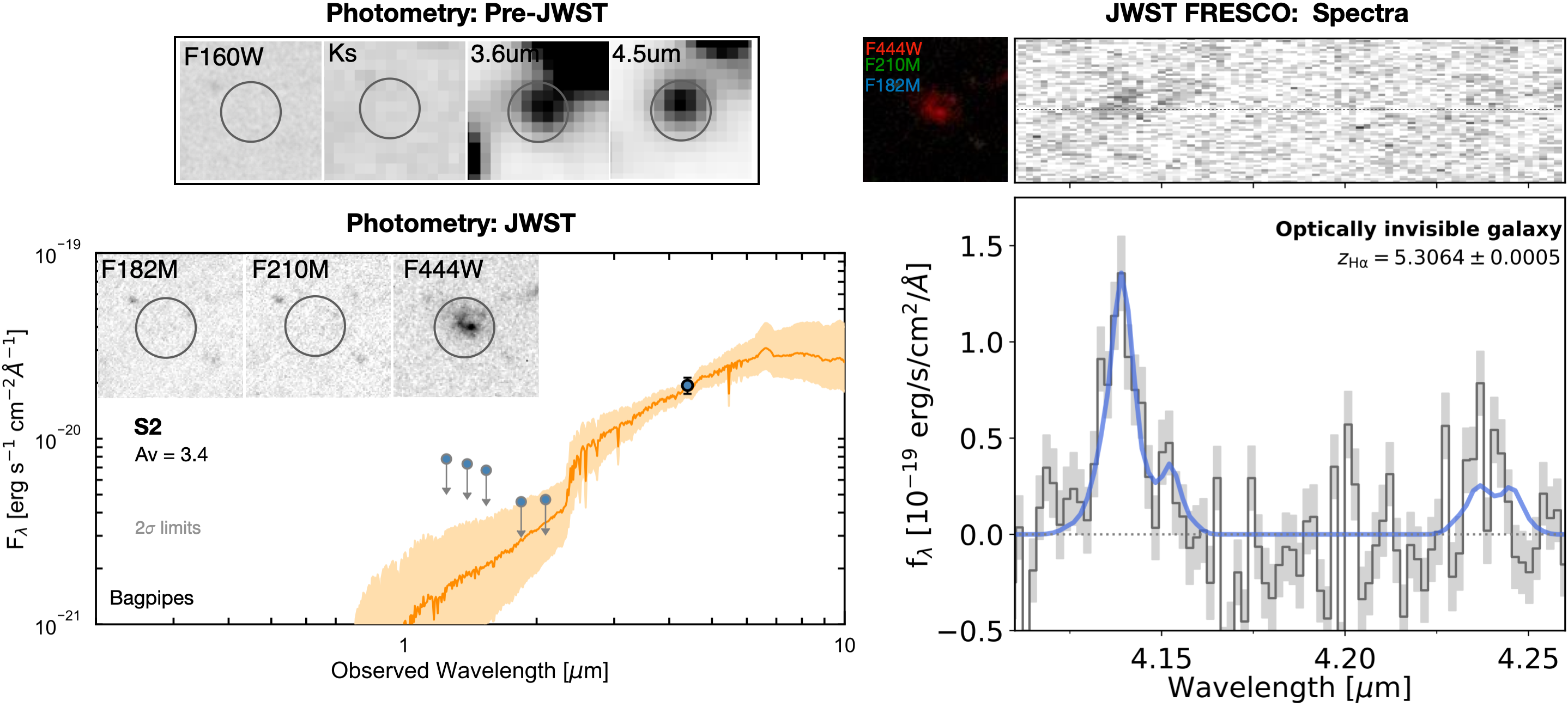} 
	\caption{\small{\textbf{Same as Extended Data Fig.~\ref{Extended_S1} but for the S2.}  
    }}
	\label{Extended_S2}
\end{Extended Data Figure}

\begin{Extended Data Figure}
	\centering
	\includegraphics[width=1.01\textwidth]{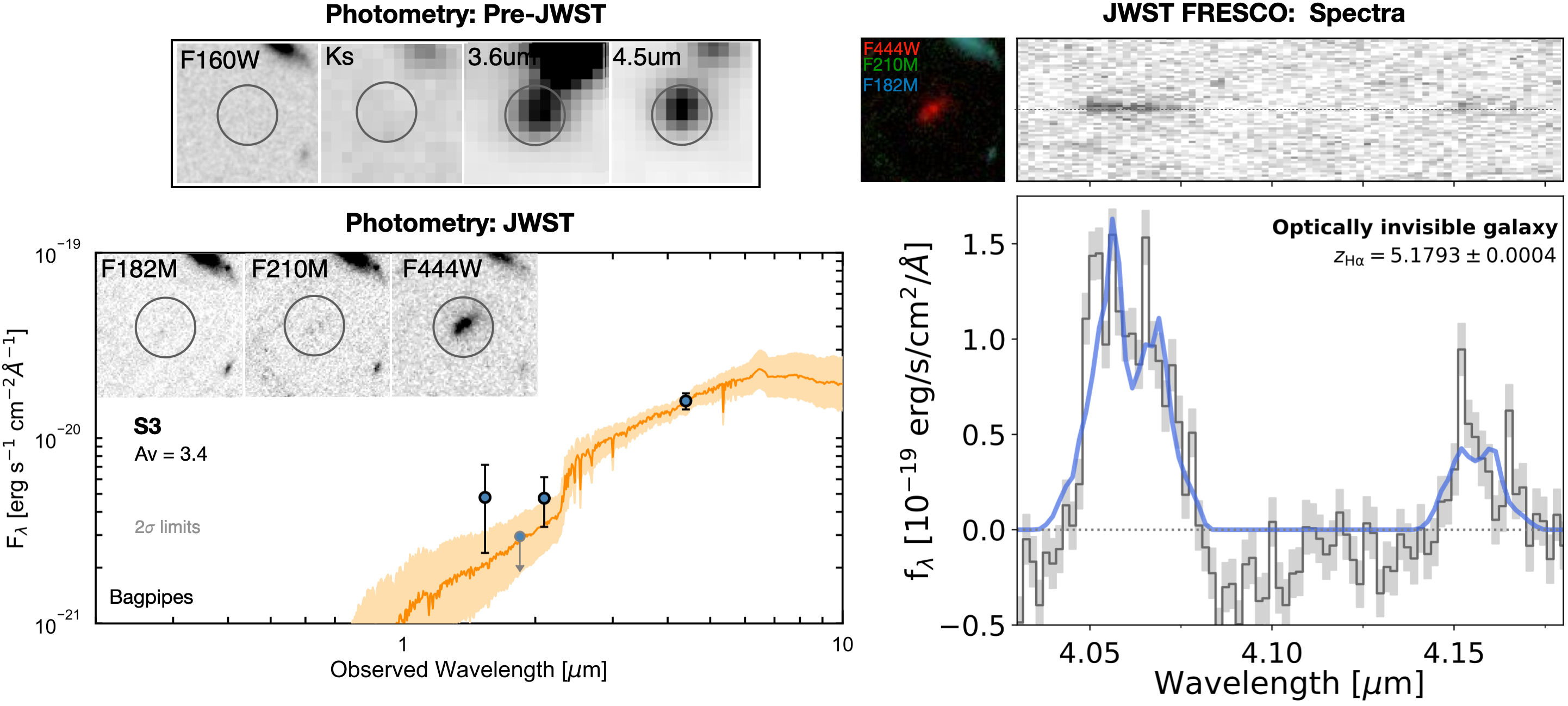} 
	\caption{\small{\textbf{Same as Extended Data Fig.~\ref{Extended_S1} but for the S3.}  
    }}
	\label{Extended_S3}
\end{Extended Data Figure}

\begin{Extended Data Figure}
	\centering
	\includegraphics[width=1.01\textwidth]{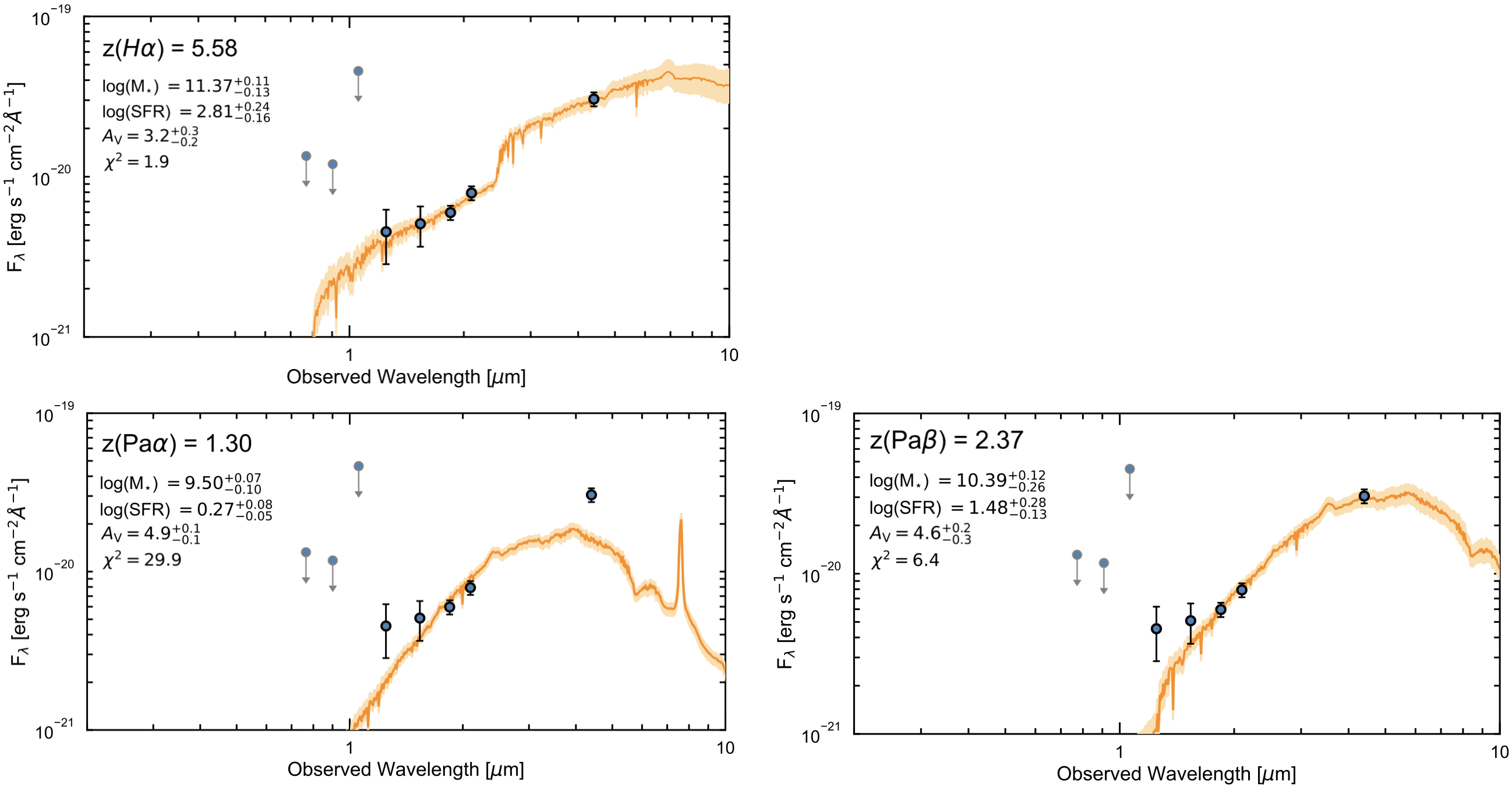} 
	\caption{\small{\textbf{Different redshift solution for the S1.}  Since we detect only one emission line ($\sim14\sigma$) in S1, which could be the H$\alpha$, Pa$\alpha$, and Pa$\beta$ lines. We show SED fitting results from \texttt{Bagpipes} with the fixed $z_{\rm spec}$ corresponding to the solution of different emission lines under the same parameter settings, respectively. Error bars correspond to 1$\sigma$ uncertainties, and downward arrows represent 2$\sigma$ upper limits. 
 The solution of the H$\alpha$ line has the least chi-square $\chi^2$ (see more discussion in Methods).
    }}
	\label{solution}
\end{Extended Data Figure}

\begin{Extended Data Figure}
	\centering
	\includegraphics[width=\textwidth]{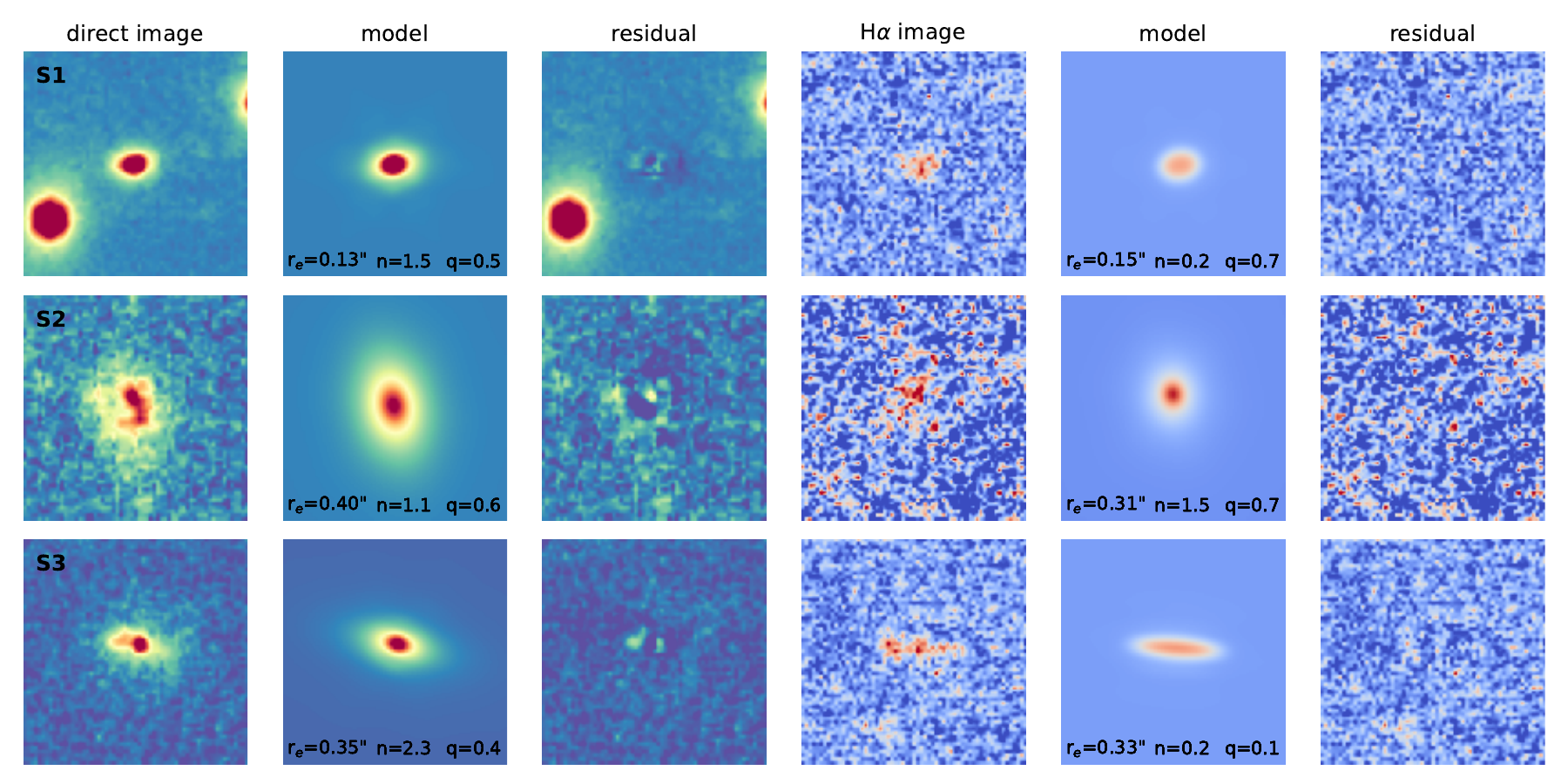} 
	\caption{\small{\textbf{Structural parameter fits.} We show our inference of the structural parameters of each source for both the F444W direct image and the H$\alpha$ map from the grism spectrum. The extended morphologies of the sources in both the direct and H$\alpha$ images suggest that the light from these objects in not dominated by AGN. 
    }}
	\label{galfit}
\end{Extended Data Figure}

% to use a second bibliography specifically for the appendix, use bibunit.
% need to run bibtext separately though and edit bu2.bbl. See instructions below

\end{bibunit}


\begin{thebibliography}{10}
\small
\expandafter\ifx\csname url\endcsname\relax
  \def\url#1{\texttt{#1}}\fi
\expandafter\ifx\csname urlprefix\endcsname\relax\def\urlprefix{URL }\fi
\providecommand{\bibinfo}[2]{#2}
\providecommand{\eprint}[2][]{\url{#2}}

\bibitem{Smail1997} 
 \bibinfo{author}{Smail, Ian, Ivison, R. J., \& Blain, A. W.}  
 \newblock \bibinfo{title}{{A Deep Sub-millimeter Survey of Lensing Clusters: A New Window on Galaxy Formation and Evolution}}. 
 \newblock \emph{\bibinfo{journal}{The Astrophysical Journal}} 
 \textbf{\bibinfo{volume}{490}}, \bibinfo{pages}{L5-L8} 
 (\bibinfo{year}{1997}). 

\bibitem{Hughes1998} 
 \bibinfo{author}{Hughes, David H., et al.}  
 \newblock \bibinfo{title}{{High-redshift star formation in the Hubble Deep Field revealed by a submillimetre-wavelength survey}}. 
 \newblock \emph{\bibinfo{journal}{Nature}} 
 \textbf{\bibinfo{volume}{394}}, \bibinfo{pages}{241-247} 
 (\bibinfo{year}{1998}). 

\bibitem{Smail2004} 
 \bibinfo{author}{Smail, Ian, et al.}  
 \newblock \bibinfo{title}{{The Rest-Frame Optical Properties of SCUBA Galaxies}}. 
 \newblock \emph{\bibinfo{journal}{The Astrophysical Journal}} 
 \textbf{\bibinfo{volume}{616}}, \bibinfo{pages}{71-85} 
 (\bibinfo{year}{2004}). 

\bibitem{Walter2012} 
 \bibinfo{author}{Walter, Fabian, et al.}  
 \newblock \bibinfo{title}{{The intense starburst HDF850.1 in a galaxy overdensity at $z \approx 5.2$ in the Hubble Deep Field}}. 
 \newblock \emph{\bibinfo{journal}{Nature}} 
 \textbf{\bibinfo{volume}{486}}, \bibinfo{pages}{233-236} 
 (\bibinfo{year}{2012}). 

\bibitem{Wang2019} 
 \bibinfo{author}{Wang, T., et al.}  
 \newblock \bibinfo{title}{{A dominant population of optically invisible massive galaxies in the early Universe}}. 
 \newblock \emph{\bibinfo{journal}{Nature}} 
 \textbf{\bibinfo{volume}{572}}, \bibinfo{pages}{211-214} 
 (\bibinfo{year}{2019}). 

\bibitem{Hodge2020} 
 \bibinfo{author}{Hodge, J. A. \& da Cunha, E.}  
 \newblock \bibinfo{title}{{High-redshift star formation in the Atacama large millimetre/submillimetre array era}}. 
 \newblock \emph{\bibinfo{journal}{Royal Society Open Science}} 
 \textbf{\bibinfo{volume}{7}}, \bibinfo{pages}{200556} 
 (\bibinfo{year}{2020}). 

 
\bibitem{Menci2022} 
 \bibinfo{author}{Menci, N., et al.}  
 \newblock \bibinfo{title}{{High-redshift Galaxies from Early JWST Observations: Constraints on Dark Energy Models}}. 
 \newblock \emph{\bibinfo{journal}{The Astrophysical Journal}} 
 \textbf{\bibinfo{volume}{938}}, \bibinfo{pages}{L5} 
 (\bibinfo{year}{2022}). 

\bibitem{Boylan-Kolchin2023} 
 \bibinfo{author}{Boylan-Kolchin, Michael}  
 \newblock \bibinfo{title}{{Stress testing $\Lambda$CDM with high-redshift galaxy candidates}}. 
 \newblock \emph{\bibinfo{journal}{Nature Astronomy}} 
 \textbf{\bibinfo{volume}{7}}, \bibinfo{pages}{731-735} 
 (\bibinfo{year}{2023}). 

\bibitem{Lovell2023} 
 \bibinfo{author}{Lovell, Christopher C., et al.}  
 \newblock \bibinfo{title}{{Extreme value statistics of the halo and stellar mass distributions at high redshift: are JWST results in tension with $\Lambda$CDM?}} 
 \newblock \emph{\bibinfo{journal}{Monthly Notices of the Royal Astronomical Society}} 
 \textbf{\bibinfo{volume}{518}}, \bibinfo{pages}{2511-2520} 
 (\bibinfo{year}{2023}). 




\bibitem{Naidu2022} 
 \bibinfo{author}{Naidu, Rohan P., et al.}  
 \newblock \bibinfo{title}{{Two Remarkably Luminous Galaxy Candidates at z $\sim$ 10-12 Revealed by JWST}}. 
 \newblock \emph{\bibinfo{journal}{The Astrophysical Journal}} 
 \textbf{\bibinfo{volume}{940}}, \bibinfo{pages}{L14} 
 (\bibinfo{year}{2022}). 

\bibitem{Castellano2022} 
 \bibinfo{author}{Castellano, Marco, et al.}  
 \newblock \bibinfo{title}{{Early Results from GLASS-JWST. III. Galaxy Candidates at z $\sim$ 9-15}}. 
 \newblock \emph{\bibinfo{journal}{The Astrophysical Journal}} 
 \textbf{\bibinfo{volume}{938}}, \bibinfo{pages}{L15} 
 (\bibinfo{year}{2022}).


\bibitem{Labbé2023} 
 \bibinfo{author}{Labbé, Ivo, et al.}  
 \newblock \bibinfo{title}{{A population of red candidate massive galaxies  600 Myr after the Big Bang}}. 
 \newblock \emph{\bibinfo{journal}{Nature}} 
 \textbf{\bibinfo{volume}{616}}, \bibinfo{pages}{266-269} 
 (\bibinfo{year}{2023}). 


\bibitem{Pérez-González2023} 
 \bibinfo{author}{Pérez-González, Pablo G., et al.}  
 \newblock \bibinfo{title}{{Life beyond 30: Probing the -20 $< M_{\rm UV}<$ -17 Luminosity Function at 8 $< z <$ 13 with the NIRCam Parallel Field of the MIRI Deep Survey}}. 
 \newblock \emph{\bibinfo{journal}{The Astrophysical Journal}} 
 \textbf{\bibinfo{volume}{951}}, \bibinfo{pages}{L1} 
 (\bibinfo{year}{2023}). 


\bibitem{Finkelstein2023} 
 \bibinfo{author}{Finkelstein, Steven L., et al.}  
 \newblock \bibinfo{title}{{The Complete CEERS Early Universe Galaxy Sample: A Surprisingly Slow Evolution of the Space Density of Bright Galaxies at z $\sim$ 8.5-14.5}}. 
 \newblock \emph{\bibinfo{journal}{arXiv e-prints}} 
 \textbf{\bibinfo{volume}}, \bibinfo{pages}{arXiv:2311.04279} 
 (\bibinfo{year}{2023}). 

\bibitem{Willott2024} 
 \bibinfo{author}{Willott, Chris J., et al.}  
 \newblock \bibinfo{title}{{A Steep Decline in the Galaxy Space Density beyond Redshift 9 in the CANUCS UV Luminosity Function}}. 
 \newblock \emph{\bibinfo{journal}{The Astrophysical Journal}} 
 \textbf{\bibinfo{volume}{966}}, \bibinfo{pages}{74} 
 (\bibinfo{year}{2024}). 

\bibitem{McLeod2024} 
 \bibinfo{author}{McLeod, D. J., et al.}  
 \newblock \bibinfo{title}{{The galaxy UV luminosity function at z $\sim$ 11 from a suite of public JWST ERS, ERO, and Cycle-1 programs}}. 
 \newblock \emph{\bibinfo{journal}{Monthly Notices of the Royal Astronomical Society}} 
 \textbf{\bibinfo{volume}{527}}, \bibinfo{pages}{5004-5022} 
 (\bibinfo{year}{2024}). 

\bibitem{Madau2014} 
 \bibinfo{author}{Madau, Piero \& Dickinson, Mark}  
 \newblock \bibinfo{title}{{Cosmic Star-Formation History}}. 
 \newblock \emph{\bibinfo{journal}{Annual Review of Astronomy and Astrophysics}} 
 \textbf{\bibinfo{volume}{52}}, \bibinfo{pages}{415-486} 
 (\bibinfo{year}{2014}). 

 
\bibitem{Oesch2023} 
 \bibinfo{author}{Oesch, P. A., et al.}  
 \newblock \bibinfo{title}{{The JWST FRESCO survey: legacy NIRCam/grism spectroscopy and imaging in the two GOODS fields}}. 
 \newblock \emph{\bibinfo{journal}{Monthly Notices of the Royal Astronomical Society}} 
 \textbf{\bibinfo{volume}{525}}, \bibinfo{pages}{2864-2874} 
 (\bibinfo{year}{2023}). 


\bibitem{Giavalisco2004} 
 \bibinfo{author}{Giavalisco, M., et al.}  
 \newblock \bibinfo{title}{{The Great Observatories Origins Deep Survey: Initial Results from Optical and Near-Infrared Imaging}}. 
 \newblock \emph{\bibinfo{journal}{The Astrophysical Journal}} 
 \textbf{\bibinfo{volume}{600}}, \bibinfo{pages}{L93-L98} 
 (\bibinfo{year}{2004}).  

\bibitem{Fudamoto2021} 
 \bibinfo{author}{Fudamoto, Y., et al.}  
 \newblock \bibinfo{title}{{Normal, dust-obscured galaxies in the epoch of reionization}}. 
 \newblock \emph{\bibinfo{journal}{Nature}} 
 \textbf{\bibinfo{volume}{597}}, \bibinfo{pages}{489-492} 
 (\bibinfo{year}{2021}). 

\bibitem{Xiao2023} 
 \bibinfo{author}{Xiao, M. -Y., et al.}  
 \newblock \bibinfo{title}{{The hidden side of cosmic star formation at $z>3$. Bridging optically dark and Lyman-break galaxies with GOODS-ALMA}}. 
 \newblock \emph{\bibinfo{journal}{Astronomy and Astrophysics}} 
 \textbf{\bibinfo{volume}{672}}, \bibinfo{pages}{A18} 
 (\bibinfo{year}{2023}).

\bibitem{Barrufet2023} 
 \bibinfo{author}{Barrufet, L., et al.}  
 \newblock \bibinfo{title}{{Unveiling the nature of infrared bright, optically dark galaxies with early JWST data}}. 
 \newblock \emph{\bibinfo{journal}{Monthly Notices of the Royal Astronomical Society}} 
 \textbf{\bibinfo{volume}{522}}, \bibinfo{pages}{449-456} 
 (\bibinfo{year}{2023}). 


\bibitem{Pérez-González2023b} 
 \bibinfo{author}{Pérez-González, Pablo G., et al.}  
 \newblock \bibinfo{title}{{CEERS Key Paper. IV. A Triality in the Nature of HST-dark Galaxies}}. 
 \newblock \emph{\bibinfo{journal}{The Astrophysical Journal}} 
 \textbf{\bibinfo{volume}{946}}, \bibinfo{pages}{L16} 
 (\bibinfo{year}{2023}). 



\bibitem{Whitaker2019} 
 \bibinfo{author}{Whitaker, Katherine E., et al.}  
 \newblock \bibinfo{title}{{The Hubble Legacy Field GOODS-S Photometric Catalog}}. 
 \newblock \emph{\bibinfo{journal}{The Astrophysical Journal Supplement Series}} 
 \textbf{\bibinfo{volume}{244}}, \bibinfo{pages}{16} 
 (\bibinfo{year}{2019}). 

\bibitem{Rieke2023} 
 \bibinfo{author}{Rieke, Marcia J., et al.}  
 \newblock \bibinfo{title}{{JADES Initial Data Release for the Hubble Ultra Deep Field: Revealing the Faint Infrared Sky with Deep JWST NIRCam Imaging}}. 
 \newblock \emph{\bibinfo{journal}{The Astrophysical Journal Supplement Series}} 
 \textbf{\bibinfo{volume}{269}}, \bibinfo{pages}{16} 
 (\bibinfo{year}{2023}). 

\bibitem{Carnall2018} 
 \bibinfo{author}{Carnall, A. C., et al.}  
 \newblock \bibinfo{title}{{Inferring the star formation histories of massive quiescent galaxies with BAGPIPES: evidence for multiple quenching mechanisms}}. 
 \newblock \emph{\bibinfo{journal}{Monthly Notices of the Royal Astronomical Society}} 
 \textbf{\bibinfo{volume}{480}}, \bibinfo{pages}{4379-4401} 
 (\bibinfo{year}{2018}). 

\bibitem{Boquien2019} 
 \bibinfo{author}{Boquien, M., et al.}  
 \newblock \bibinfo{title}{{CIGALE: a python Code Investigating GALaxy Emission}}. 
 \newblock \emph{\bibinfo{journal}{Astronomy and Astrophysics}} 
 \textbf{\bibinfo{volume}{622}}, \bibinfo{pages}{A103} 
 (\bibinfo{year}{2019}). 

\bibitem{Planck2020} 
 \bibinfo{author}{Planck Collaboration, et al.}  
 \newblock \bibinfo{title}{{Planck 2018 results. VI. Cosmological parameters}}. 
 \newblock \emph{\bibinfo{journal}{Astronomy and Astrophysics}} 
 \textbf{\bibinfo{volume}{641}}, \bibinfo{pages}{A6} 
 (\bibinfo{year}{2020}). 






\bibitem{Moster2013} 
 \bibinfo{author}{Moster, Benjamin P., Naab, Thorsten, \& White, Simon D. M.}  
 \newblock \bibinfo{title}{{Galactic star formation and accretion histories from matching galaxies to dark matter haloes}}. 
 \newblock \emph{\bibinfo{journal}{Monthly Notices of the Royal Astronomical Society}} 
 \textbf{\bibinfo{volume}{428}}, \bibinfo{pages}{3121-3138} 
 (\bibinfo{year}{2013}). 

\bibitem{Moster2018} 
 \bibinfo{author}{Moster, Benjamin P., Naab, Thorsten, \& White, Simon D. M.}  
 \newblock \bibinfo{title}{{EMERGE - an empirical model for the formation of galaxies since z $\sim$ 10}}. 
 \newblock \emph{\bibinfo{journal}{Monthly Notices of the Royal Astronomical Society}} 
 \textbf{\bibinfo{volume}{477}}, \bibinfo{pages}{1822-1852} 
 (\bibinfo{year}{2018}). 

\bibitem{Tacchella2018} 
 \bibinfo{author}{Tacchella, Sandro, et al.}  
 \newblock \bibinfo{title}{{A Redshift-independent Efficiency Model: Star Formation and Stellar Masses in Dark Matter Halos at z $\gtrsim$ 4}}. 
 \newblock \emph{\bibinfo{journal}{The Astrophysical Journal}} 
 \textbf{\bibinfo{volume}{868}}, \bibinfo{pages}{92} 
 (\bibinfo{year}{2018}). 


\bibitem{Pillepich2018} 
 \bibinfo{author}{Pillepich, Annalisa, et al.}  
 \newblock \bibinfo{title}{{Simulating galaxy formation with the IllustrisTNG model}}. 
 \newblock \emph{\bibinfo{journal}{Monthly Notices of the Royal Astronomical Society}} 
 \textbf{\bibinfo{volume}{473}}, \bibinfo{pages}{4077-4106} 
 (\bibinfo{year}{2018}). 


\bibitem{Wechsler2018} 
 \bibinfo{author}{Wechsler, Risa H. \& Tinker, Jeremy L.}  
 \newblock \bibinfo{title}{{The Connection Between Galaxies and Their Dark Matter Halos}}. 
 \newblock \emph{\bibinfo{journal}{Annual Review of Astronomy and Astrophysics}} 
 \textbf{\bibinfo{volume}{56}}, \bibinfo{pages}{435-487} 
 (\bibinfo{year}{2018}). 

\bibitem{Shuntov2022} 
 \bibinfo{author}{Shuntov, M., et al.}  
 \newblock \bibinfo{title}{{COSMOS2020: Cosmic evolution of the stellar-to-halo mass relation for central and satellite galaxies up to z $\sim$ 5}}. 
 \newblock \emph{\bibinfo{journal}{Astronomy and Astrophysics}} 
 \textbf{\bibinfo{volume}{664}}, \bibinfo{pages}{A61} 
 (\bibinfo{year}{2022}). 

\bibitem{Riechers2020} 
 \bibinfo{author}{Riechers, Dominik A., et al.}  
 \newblock \bibinfo{title}{{COLDz: A High Space Density of Massive Dusty Starburst Galaxies $\sim$1 Billion Years after the Big Bang}}. 
 \newblock \emph{\bibinfo{journal}{The Astrophysical Journal}} 
 \textbf{\bibinfo{volume}{895}}, \bibinfo{pages}{81} 
 (\bibinfo{year}{2020}). 

  
\bibitem{White1978} 
 \bibinfo{author}{White, S. D. M. \& Rees, M. J.}  
 \newblock \bibinfo{title}{{Core condensation in heavy halos: a two-stage theory for galaxy formation and clustering.}}. 
 \newblock \emph{\bibinfo{journal}{Monthly Notices of the Royal Astronomical Society}} 
 \textbf{\bibinfo{volume}{183}}, \bibinfo{pages}{341-358} 
 (\bibinfo{year}{1978}). 

 \bibitem{Dekel2023} 
 \bibinfo{author}{Dekel, Avishai, et al.}  
 \newblock \bibinfo{title}{{Efficient formation of massive galaxies at cosmic dawn by feedback-free starbursts}}. 
 \newblock \emph{\bibinfo{journal}{Monthly Notices of the Royal Astronomical Society}} 
 \textbf{\bibinfo{volume}{523}}, \bibinfo{pages}{3201-3218} 
 (\bibinfo{year}{2023}). 
 
\bibitem{Li2023} 
 \bibinfo{author}{Li, Zhaozhou, et al.}  
 \newblock \bibinfo{title}{{Feedback-Free Starbursts at Cosmic Dawn: Observable Predictions for JWST}}. 
 \newblock \emph{\bibinfo{journal}{arXiv e-prints}} 
 \textbf{\bibinfo{volume}}, \bibinfo{pages}{arXiv:2311.14662} 
 (\bibinfo{year}{2023}). 


\bibitem{Herard-Demanche2023} 
 \bibinfo{author}{Herard-Demanche, Thomas, et al.}  
 \newblock \bibinfo{title}{{Mapping dusty galaxy growth at $z>5$ with FRESCO: Detection of H$\alpha$ in submm galaxy HDF850.1 and the surrounding overdense structures}}. 
 \newblock \emph{\bibinfo{journal}{arXiv e-prints}} 
 \textbf{\bibinfo{volume}}, \bibinfo{pages}{arXiv:2309.04525} 
 (\bibinfo{year}{2023}). 


\bibitem{Schaye2023} 
 \bibinfo{author}{Schaye, Joop, et al.}  
 \newblock \bibinfo{title}{{The FLAMINGO project: cosmological hydrodynamical simulations for large-scale structure and galaxy cluster surveys}}. 
 \newblock \emph{\bibinfo{journal}{Monthly Notices of the Royal Astronomical Society}} 
 \textbf{\bibinfo{volume}{526}}, \bibinfo{pages}{4978-5020} 
 (\bibinfo{year}{2023}). 
  

\bibitem{Bouwens2012a} 
 \bibinfo{author}{Bouwens, R. J., et al.}  
 \newblock \bibinfo{title}{{UV-continuum Slopes at z $\sim$ 4-7 from the HUDF09+ERS+CANDELS Observations: Discovery of a Well-defined UV Color-Magnitude Relationship for z $>$= 4 Star-forming Galaxies}}. 
 \newblock \emph{\bibinfo{journal}{The Astrophysical Journal}} 
 \textbf{\bibinfo{volume}{754}}, \bibinfo{pages}{83} 
 (\bibinfo{year}{2012}). 

\bibitem{Bouwens2012b} 
 \bibinfo{author}{Bouwens, R. J., et al.}  
 \newblock \bibinfo{title}{{Lower-luminosity Galaxies Could Reionize the Universe: Very Steep Faint-end Slopes to the UV Luminosity Functions at z $>$= 5-8 from the HUDF09 WFC3/IR Observations}}. 
 \newblock \emph{\bibinfo{journal}{The Astrophysical Journal}} 
 \textbf{\bibinfo{volume}{752}}, \bibinfo{pages}{L5} 
 (\bibinfo{year}{2012}). 
 










\end{thebibliography}


\begin{thebibliography}{10}
\small
\expandafter\ifx\csname url\endcsname\relax
  \def\url#1{\texttt{#1}}\fi
\expandafter\ifx\csname urlprefix\endcsname\relax\def\urlprefix{URL }\fi
\providecommand{\bibinfo}[2]{#2}
\providecommand{\eprint}[2][]{\url{#2}}


 \bibitem{Chabrier2003}
\bibinfo{author}{{Chabrier}, G.}
\newblock \bibinfo{title}{{Galactic Stellar and Substellar Initial Mass Function}}.
\newblock \emph{\bibinfo{journal}{Publ. Astron. Soc. Pac.}}
  \textbf{\bibinfo{volume}{809}}, \bibinfo{pages}{763-795}
  (\bibinfo{year}{2003}).

 \bibitem{Salpeter1955}
\bibinfo{author}{{Salpeter}, E.}
\newblock \bibinfo{title}{{The Luminosity Function and Stellar Evolution}}.
\newblock \emph{\bibinfo{journal}{Astrophys. J.}}
  \textbf{\bibinfo{volume}{121}}, \bibinfo{pages}{161} (\bibinfo{year}{1955}).


\bibitem{Oke1983} 
 \bibinfo{author}{Oke, J. B. \& Gunn, J. E.}  
 \newblock \bibinfo{title}{{Secondary standard stars for absolute spectrophotometry.}}. 
 \newblock \emph{\bibinfo{journal}{The Astrophysical Journal}} 
 \textbf{\bibinfo{volume}{266}}, \bibinfo{pages}{713-717} 
 (\bibinfo{year}{1983}). 

\bibitem{Kashino2022} 
 \bibinfo{author}{Kashino, Daichi, et al.}  
 \newblock \bibinfo{title}{{The Stellar Mass versus Stellar Metallicity Relation of Star-forming Galaxies at $1.6 \leq z\leq 3.0$ and Implications for the Evolution of the $\alpha$-enhancement}}. 
 \newblock \emph{\bibinfo{journal}{The Astrophysical Journal}} 
 \textbf{\bibinfo{volume}{925}}, \bibinfo{pages}{82} 
 (\bibinfo{year}{2022}). 

\bibitem{Giavalisco1996} 
 \bibinfo{author}{Giavalisco, Mauro, Steidel, Charles C., \& Macchetto, F. Duccio}  
 \newblock \bibinfo{title}{{Hubble Space Telescope Imaging of Star-forming Galaxies at Redshifts Z $>$ 3}}. 
 \newblock \emph{\bibinfo{journal}{The Astrophysical Journal}} 
 \textbf{\bibinfo{volume}{470}}, \bibinfo{pages}{189} 
 (\bibinfo{year}{1996}). 

\bibitem{Koekemoer2011} 
 \bibinfo{author}{Koekemoer, Anton M., et al.}  
 \newblock \bibinfo{title}{{CANDELS: The Cosmic Assembly Near-infrared Deep Extragalactic Legacy Survey—The Hubble Space Telescope Observations, Imaging Data Products, and Mosaics}}. 
 \newblock \emph{\bibinfo{journal}{The Astrophysical Journal Supplement Series}} 
 \textbf{\bibinfo{volume}{197}}, \bibinfo{pages}{36} 
 (\bibinfo{year}{2011}). 

\bibitem{Grogin2011} 
 \bibinfo{author}{Grogin, Norman A., et al.}  
 \newblock \bibinfo{title}{{CANDELS: The Cosmic Assembly Near-infrared Deep Extragalactic Legacy Survey}}. 
 \newblock \emph{\bibinfo{journal}{The Astrophysical Journal Supplement Series}} 
 \textbf{\bibinfo{volume}{197}}, \bibinfo{pages}{35} 
 (\bibinfo{year}{2011}). 

\bibitem{Illingworth2016} 
 \bibinfo{author}{Illingworth, Garth, et al.}  
 \newblock \bibinfo{title}{{The Hubble Legacy Fields (HLF-GOODS-S) v1.5 Data Products: Combining 2442 Orbits of GOODS-S/CDF-S Region ACS and WFC3/IR Images}}. 
 \newblock \emph{\bibinfo{journal}{arXiv e-prints}} 
 \textbf{\bibinfo{volume}}, \bibinfo{pages}{arXiv:1606.00841} 
 (\bibinfo{year}{2016}).

\bibitem{Beckwith2006} 
 \bibinfo{author}{Beckwith, Steven V. W., et al.}  
 \newblock \bibinfo{title}{{The Hubble Ultra Deep Field}}. 
 \newblock \emph{\bibinfo{journal}{The Astronomical Journal}} 
 \textbf{\bibinfo{volume}{132}}, \bibinfo{pages}{1729-1755} 
 (\bibinfo{year}{2006}). 

\bibitem{Eisenstein2023} 
 \bibinfo{author}{Eisenstein, Daniel J., et al.}  
 \newblock \bibinfo{title}{{Overview of the JWST Advanced Deep Extragalactic Survey (JADES)}}. 
 \newblock \emph{\bibinfo{journal}{arXiv e-prints}} 
 \textbf{\bibinfo{volume}}, \bibinfo{pages}{arXiv:2306.02465} 
 (\bibinfo{year}{2023}). 

\bibitem{Williams2023} 
 \bibinfo{author}{Williams, Christina C., et al.}  
 \newblock \bibinfo{title}{{JEMS: A Deep Medium-band Imaging Survey in the Hubble Ultra Deep Field with JWST NIRCam and NIRISS}}. 
 \newblock \emph{\bibinfo{journal}{The Astrophysical Journal Supplement Series}} 
 \textbf{\bibinfo{volume}{268}}, \bibinfo{pages}{64} 
 (\bibinfo{year}{2023}). 

\bibitem{Bertin1996} 
 \bibinfo{author}{Bertin, E. \& Arnouts, S.}  
 \newblock \bibinfo{title}{{SExtractor: Software for source extraction.}}. 
 \newblock \emph{\bibinfo{journal}{Astronomy and Astrophysics Supplement Series}} 
 \textbf{\bibinfo{volume}{117}}, \bibinfo{pages}{393-404} 
 (\bibinfo{year}{1996}). 

\bibitem{Brammer2008} 
 \bibinfo{author}{Brammer, Gabriel B., van Dokkum, Pieter G., \& Coppi, Paolo}  
 \newblock \bibinfo{title}{{EAZY: A Fast, Public Photometric Redshift Code}}. 
 \newblock \emph{\bibinfo{journal}{The Astrophysical Journal}} 
 \textbf{\bibinfo{volume}{686}}, \bibinfo{pages}{1503-1513} 
 (\bibinfo{year}{2008}).

\bibitem{Weibel2024} 
 \bibinfo{author}{Weibel, Andrea, et al.}  
 \newblock \bibinfo{title}{{Galaxy Build-up in the first 1.5 Gyr of Cosmic History: Insights from the Stellar Mass Function at $z\sim4-9$ from JWST NIRCam Observations}}. 
 \newblock \emph{\bibinfo{journal}{arXiv e-prints}} 
 \textbf{\bibinfo{volume}}, \bibinfo{pages}{arXiv:2403.08872} 
 (\bibinfo{year}{2024}). 

\bibitem{Franco2018} 
 \bibinfo{author}{Franco, M., et al.}  
 \newblock \bibinfo{title}{{GOODS-ALMA: 1.1 mm galaxy survey. I. Source catalog and optically dark galaxies}}. 
 \newblock \emph{\bibinfo{journal}{Astronomy and Astrophysics}} 
 \textbf{\bibinfo{volume}{620}}, \bibinfo{pages}{A152} 
 (\bibinfo{year}{2018}). 

 \bibitem{Gómez-Guijarro2022} 
 \bibinfo{author}{Gómez-Guijarro, C., et al.}  
 \newblock \bibinfo{title}{{GOODS-ALMA 2.0: Source catalog, number counts, and prevailing compact sizes in 1.1 mm galaxies}}. 
 \newblock \emph{\bibinfo{journal}{Astronomy and Astrophysics}} 
 \textbf{\bibinfo{volume}{658}}, \bibinfo{pages}{A43} 
 (\bibinfo{year}{2022}).

\bibitem{Cowie2018} 
 \bibinfo{author}{Cowie, L. L., et al.}  
 \newblock \bibinfo{title}{{A Submillimeter Perspective on the GOODS Fields (SUPER GOODS). III. A Large Sample of ALMA Sources in the GOODS-S}}. 
 \newblock \emph{\bibinfo{journal}{The Astrophysical Journal}} 
 \textbf{\bibinfo{volume}{865}}, \bibinfo{pages}{106} 
 (\bibinfo{year}{2018}). 

\bibitem{Cowie2017} 
 \bibinfo{author}{Cowie, L. L., et al.}  
 \newblock \bibinfo{title}{{A Submillimeter Perspective on the GOODS Fields (SUPER GOODS). I. An Ultradeep SCUBA-2 Survey of the GOODS-N}}. 
 \newblock \emph{\bibinfo{journal}{The Astrophysical Journal}} 
 \textbf{\bibinfo{volume}{837}}, \bibinfo{pages}{139} 
 (\bibinfo{year}{2017}). 

\bibitem{Barger2022} 
 \bibinfo{author}{Barger, A. J., et al.}  
 \newblock \bibinfo{title}{{A Submillimeter Perspective on the GOODS Fields (SUPER GOODS). V. Deep 450 $\mu$m Imaging}}. 
 \newblock \emph{\bibinfo{journal}{The Astrophysical Journal}} 
 \textbf{\bibinfo{volume}{934}}, \bibinfo{pages}{56} 
 (\bibinfo{year}{2022}). 

\bibitem{Alcalde2019} 
 \bibinfo{author}{Alcalde Pampliega, Belén, et al.}  
 \newblock \bibinfo{title}{{Optically Faint Massive Balmer Break Galaxies at $z > 3$ in the CANDELS/GOODS Fields}}. 
 \newblock \emph{\bibinfo{journal}{The Astrophysical Journal}} 
 \textbf{\bibinfo{volume}{876}}, \bibinfo{pages}{135} 
 (\bibinfo{year}{2019}). 


\bibitem{Gómez-Guijarro2023} 
 \bibinfo{author}{Gómez-Guijarro, Carlos, et al.}  
 \newblock \bibinfo{title}{{JWST CEERS probes the role of stellar mass and morphology in obscuring galaxies}}. 
 \newblock \emph{\bibinfo{journal}{Astronomy and Astrophysics}} 
 \textbf{\bibinfo{volume}{677}}, \bibinfo{pages}{A34} 
 (\bibinfo{year}{2023}). 

\bibitem{McKinney2023} 
 \bibinfo{author}{McKinney, Jed, et al.}  
 \newblock \bibinfo{title}{{A Near-infrared-faint, Far-infrared-luminous Dusty Galaxy at $z\sim5$ in COSMOS-Web}}. 
 \newblock \emph{\bibinfo{journal}{The Astrophysical Journal}} 
 \textbf{\bibinfo{volume}{956}}, \bibinfo{pages}{72} 
 (\bibinfo{year}{2023}).  

\bibitem{Barro2024} 
 \bibinfo{author}{Barro, Guillermo, et al.}  
 \newblock \bibinfo{title}{{Extremely Red Galaxies at z = 5–9 with MIRI and NIRSpec: Dusty Galaxies or Obscured Active Galactic Nuclei?}}. 
 \newblock \emph{\bibinfo{journal}{The Astrophysical Journal}} 
 \textbf{\bibinfo{volume}{963}}, \bibinfo{pages}{128} 
 (\bibinfo{year}{2024}). 

\bibitem{Akins2023} 
 \bibinfo{author}{Akins, Hollis B., et al.}  
 \newblock \bibinfo{title}{{Two Massive, Compact, and Dust-obscured Candidate $z \sim 8$ Galaxies Discovered by JWST}}. 
 \newblock \emph{\bibinfo{journal}{The Astrophysical Journal}} 
 \textbf{\bibinfo{volume}{956}}, \bibinfo{pages}{61} 
 (\bibinfo{year}{2023}). 

\bibitem{Williams2019} 
 \bibinfo{author}{Williams, Christina C., et al.}  
 \newblock \bibinfo{title}{{Discovery of a Dark, Massive, ALMA-only Galaxy at $z \sim 5-6$ in a Tiny 3 mm Survey}}. 
 \newblock \emph{\bibinfo{journal}{The Astrophysical Journal}} 
 \textbf{\bibinfo{volume}{884}}, \bibinfo{pages}{154} 
 (\bibinfo{year}{2019}). 

\bibitem{vanderVlugt2023} 
 \bibinfo{author}{van der Vlugt, D., et al.}  
 \newblock \bibinfo{title}{{An Ultradeep Multiband Very Large Array Survey of the Faint Radio Sky (COSMOS-XS): New Constraints on the Optically Dark Population}}. 
 \newblock \emph{\bibinfo{journal}{The Astrophysical Journal}} 
 \textbf{\bibinfo{volume}{951}}, \bibinfo{pages}{131} 
 (\bibinfo{year}{2023}). 


\bibitem{Bruzual2003} 
 \bibinfo{author}{Bruzual, G. \& Charlot, S.}  
 \newblock \bibinfo{title}{{Stellar population synthesis at the resolution of 2003}}. 
 \newblock \emph{\bibinfo{journal}{Monthly Notices of the Royal Astronomical Society}} 
 \textbf{\bibinfo{volume}{344}}, \bibinfo{pages}{1000-1028} 
 (\bibinfo{year}{2003}). 

\bibitem{Calzetti2000} 
 \bibinfo{author}{Calzetti, Daniela, et al.}  
 \newblock \bibinfo{title}{{The Dust Content and Opacity of Actively Star-forming Galaxies}}. 
 \newblock \emph{\bibinfo{journal}{The Astrophysical Journal}} 
 \textbf{\bibinfo{volume}{533}}, \bibinfo{pages}{682-695} 
 (\bibinfo{year}{2000}). 

\bibitem{Elbaz2018} 
 \bibinfo{author}{Elbaz, D., et al.}  
 \newblock \bibinfo{title}{{Starbursts in and out of the star-formation main sequence}}. 
 \newblock \emph{\bibinfo{journal}{Astronomy and Astrophysics}} 
 \textbf{\bibinfo{volume}{616}}, \bibinfo{pages}{A110} 
 (\bibinfo{year}{2018}). 

\bibitem{Puglisi2017} 
 \bibinfo{author}{Puglisi, A., et al.}  
 \newblock \bibinfo{title}{{The Bright and Dark Sides of High-redshift Starburst Galaxies from Herschel and Subaru Observations}}. 
 \newblock \emph{\bibinfo{journal}{The Astrophysical Journal}} 
 \textbf{\bibinfo{volume}{838}}, \bibinfo{pages}{L18} 
 (\bibinfo{year}{2017}). 

\bibitem{Matthee2024} 
 \bibinfo{author}{Matthee, Jorryt, et al.}  
 \newblock \bibinfo{title}{{Little Red Dots: An Abundant Population of Faint Active Galactic Nuclei at $z \sim 5$ Revealed by the EIGER and FRESCO JWST Surveys}}. 
 \newblock \emph{\bibinfo{journal}{The Astrophysical Journal}} 
 \textbf{\bibinfo{volume}{963}}, \bibinfo{pages}{129} 
 (\bibinfo{year}{2024}). 


\bibitem{Landt2015} 
 \bibinfo{author}{Landt, Hermine, et al.}  
 \newblock \bibinfo{title}{{Variability of the coronal line region in NGC 4151}}. 
 \newblock \emph{\bibinfo{journal}{Monthly Notices of the Royal Astronomical Society}} 
 \textbf{\bibinfo{volume}{449}}, \bibinfo{pages}{3795-3805} 
 (\bibinfo{year}{2015}). 

\bibitem{Charlot2000} 
 \bibinfo{author}{Charlot, Stéphane \& Fall, S. Michael}  
 \newblock \bibinfo{title}{{A Simple Model for the Absorption of Starlight by Dust in Galaxies}}. 
 \newblock \emph{\bibinfo{journal}{The Astrophysical Journal}} 
 \textbf{\bibinfo{volume}{539}}, \bibinfo{pages}{718-731} 
 (\bibinfo{year}{2000}). 

\bibitem{Salim2018} 
 \bibinfo{author}{Salim, Samir, Boquien, Médéric, \& Lee, Janice C.}  
 \newblock \bibinfo{title}{{Dust Attenuation Curves in the Local Universe: Demographics and New Laws for Star-forming Galaxies and High-redshift Analogs}}. 
 \newblock \emph{\bibinfo{journal}{The Astrophysical Journal}} 
 \textbf{\bibinfo{volume}{859}}, \bibinfo{pages}{11} 
 (\bibinfo{year}{2018}). 

\bibitem{Schreiber2018} 
 \bibinfo{author}{Schreiber, C., et al.}  
 \newblock \bibinfo{title}{{Dust temperature and mid-to-total infrared color distributions for star-forming galaxies at $0 < z < 4$}}. 
 \newblock \emph{\bibinfo{journal}{Astronomy and Astrophysics}} 
 \textbf{\bibinfo{volume}{609}}, \bibinfo{pages}{A30} 
 (\bibinfo{year}{2018}). 

\bibitem{Reddy2023} 
 \bibinfo{author}{Reddy, Naveen A., et al.}  
 \newblock \bibinfo{title}{{Paschen-line Constraints on Dust Attenuation and Star Formation at z   1-3 with JWST/NIRSpec}}. 
 \newblock \emph{\bibinfo{journal}{The Astrophysical Journal}} 
 \textbf{\bibinfo{volume}{948}}, \bibinfo{pages}{83} 
 (\bibinfo{year}{2023}). 


\bibitem{Zhou2020} 
 \bibinfo{author}{Zhou, L., et al.}  
 \newblock \bibinfo{title}{{GOODS-ALMA: Optically dark ALMA galaxies shed light on a cluster in formation at z = 3.5}}. 
 \newblock \emph{\bibinfo{journal}{Astronomy and Astrophysics}} 
 \textbf{\bibinfo{volume}{642}}, \bibinfo{pages}{A155} 
 (\bibinfo{year}{2020}). 

\bibitem{Jin2022} 
 \bibinfo{author}{Jin, Shuowen, et al.}  
 \newblock \bibinfo{title}{{Diagnosing deceivingly cold dusty galaxies at $3.5 < z < 6$: A substantial population of compact starbursts with high infrared optical depths}}. 
 \newblock \emph{\bibinfo{journal}{Astronomy and Astrophysics}} 
 \textbf{\bibinfo{volume}{665}}, \bibinfo{pages}{A3} 
 (\bibinfo{year}{2022}). 

\bibitem{Kocevski2023} 
 \bibinfo{author}{Kocevski, Dale D., et al.}  
 \newblock \bibinfo{title}{{Hidden Little Monsters: Spectroscopic Identification of Low-mass, Broad-line AGNs at z $>$ 5 with CEERS}}. 
 \newblock \emph{\bibinfo{journal}{The Astrophysical Journal}} 
 \textbf{\bibinfo{volume}{954}}, \bibinfo{pages}{L4} 
 (\bibinfo{year}{2023}). 

\bibitem{Labbe2023b} 
 \bibinfo{author}{Labbe, Ivo, et al.}  
 \newblock \bibinfo{title}{{UNCOVER: Candidate Red Active Galactic Nuclei at 3$<z<$7 with JWST and ALMA}}. 
 \newblock \emph{\bibinfo{journal}{arXiv e-prints}} 
 \textbf{\bibinfo{volume}}, \bibinfo{pages}{arXiv:2306.07320} 
 (\bibinfo{year}{2023}). 

\bibitem{Peng2002} 
 \bibinfo{author}{Peng, Chien Y., et al.}  
 \newblock \bibinfo{title}{{Detailed Structural Decomposition of Galaxy Images}}. 
 \newblock \emph{\bibinfo{journal}{The Astronomical Journal}} 
 \textbf{\bibinfo{volume}{124}}, \bibinfo{pages}{266-293} 
 (\bibinfo{year}{2002}). 

\bibitem{Perrin2014} 
 \bibinfo{author}{Perrin, Marshall D., et al.}  
 \newblock \bibinfo{title}{{Updated point spread function simulations for JWST with WebbPSF}}. 
 \newblock \emph{\bibinfo{journal}{Space Telescopes and Instrumentation 2014: Optical, Infrared, and Millimeter Wave}} 
 \textbf{\bibinfo{volume}{9143}}, \bibinfo{pages}{91433X} 
 (\bibinfo{year}{2014}). 

\bibitem{Draine2014} 
 \bibinfo{author}{Draine, B. T., et al.}  
 \newblock \bibinfo{title}{{Andromeda's Dust}}. 
 \newblock \emph{\bibinfo{journal}{The Astrophysical Journal}} 
 \textbf{\bibinfo{volume}{780}}, \bibinfo{pages}{172} 
 (\bibinfo{year}{2014}). 

\bibitem{Kennicutt2012} 
 \bibinfo{author}{Kennicutt, Robert C. \& Evans, Neal J.}  
 \newblock \bibinfo{title}{{Star Formation in the Milky Way and Nearby Galaxies}}. 
 \newblock \emph{\bibinfo{journal}{Annual Review of Astronomy and Astrophysics}} 
 \textbf{\bibinfo{volume}{50}}, \bibinfo{pages}{531-608} 
 (\bibinfo{year}{2012}). 


\bibitem{Tinker2008} 
 \bibinfo{author}{Tinker, Jeremy, et al.}  
 \newblock \bibinfo{title}{{Toward a Halo Mass Function for Precision Cosmology: The Limits of Universality}}. 
 \newblock \emph{\bibinfo{journal}{The Astrophysical Journal}} 
 \textbf{\bibinfo{volume}{688}}, \bibinfo{pages}{709-728} 
 (\bibinfo{year}{2008}). 

\bibitem{Murray2013} 
 \bibinfo{author}{Murray, S. G., Power, C., \& Robotham, A. S. G.}  
 \newblock \bibinfo{title}{{HMFcalc: An online tool for calculating dark matter halo mass functions}}. 
 \newblock \emph{\bibinfo{journal}{Astronomy and Computing}} 
 \textbf{\bibinfo{volume}{3}}, \bibinfo{pages}{23} 
 (\bibinfo{year}{2013}). 


\bibitem{Kugel2023} 
 \bibinfo{author}{Kugel, Roi, et al.}  
 \newblock \bibinfo{title}{{FLAMINGO: calibrating large cosmological hydrodynamical simulations with machine learning}}. 
 \newblock \emph{\bibinfo{journal}{Monthly Notices of the Royal Astronomical Society}} 
 \textbf{\bibinfo{volume}{526}}, \bibinfo{pages}{6103-6127} 
 (\bibinfo{year}{2023}). 

\bibitem{Bunker2023} 
 \bibinfo{author}{Bunker, Andrew J., et al.}  
 \newblock \bibinfo{title}{{JADES NIRSpec Initial Data Release for the Hubble Ultra Deep Field: Redshifts and Line Fluxes of Distant Galaxies from the Deepest JWST Cycle 1 NIRSpec Multi-Object Spectroscopy}}. 
 \newblock \emph{\bibinfo{journal}{arXiv e-prints}} 
 \textbf{\bibinfo{volume}}, \bibinfo{pages}{arXiv:2306.02467} 
 (\bibinfo{year}{2023}). 

\bibitem{Yamaguchi2019} 
 \bibinfo{author}{Yamaguchi, Yuki, et al.}  
 \newblock \bibinfo{title}{{ALMA 26 arcmin$^{2}$ Survey of GOODS-S at 1 mm (ASAGAO): Near-infrared-dark Faint ALMA Sources}}. 
 \newblock \emph{\bibinfo{journal}{The Astrophysical Journal}} 
 \textbf{\bibinfo{volume}{878}}, \bibinfo{pages}{73} 
 (\bibinfo{year}{2019}). 

\bibitem{Hainline2024} 
 \bibinfo{author}{Hainline, Kevin N., et al.}  
 \newblock \bibinfo{title}{{The Cosmos in Its Infancy: JADES Galaxy Candidates at $z > 8$ in GOODS-S and GOODS-N}}. 
 \newblock \emph{\bibinfo{journal}{The Astrophysical Journal}} 
 \textbf{\bibinfo{volume}{964}}, \bibinfo{pages}{71} 
 (\bibinfo{year}{2024}). 

\bibitem{Schreiber2015} 
 \bibinfo{author}{Schreiber, C., et al.}  
 \newblock \bibinfo{title}{{The Herschel view of the dominant mode of galaxy growth from z = 4 to the present day}}. 
 \newblock \emph{\bibinfo{journal}{Astronomy and Astrophysics}} 
 \textbf{\bibinfo{volume}{575}}, \bibinfo{pages}{A74} 
 (\bibinfo{year}{2015}). 


\bibitem{Popesso2023} 
 \bibinfo{author}{Popesso, P., et al.}  
 \newblock \bibinfo{title}{{The main sequence of star-forming galaxies across cosmic times}}. 
 \newblock \emph{\bibinfo{journal}{Monthly Notices of the Royal Astronomical Society}} 
 \textbf{\bibinfo{volume}{519}}, \bibinfo{pages}{1526-1544} 
 (\bibinfo{year}{2023}). 


\bibitem{Straatman2016} 
 \bibinfo{author}{Straatman, Caroline M. S., et al.}  
 \newblock \bibinfo{title}{{The FourStar Galaxy Evolution Survey (ZFOURGE): Ultraviolet to Far-infrared Catalogs, Medium-bandwidth Photometric Redshifts with Improved Accuracy, Stellar Masses, and Confirmation of Quiescent Galaxies to $z\sim3.5$}}. 
 \newblock \emph{\bibinfo{journal}{The Astrophysical Journal}} 
 \textbf{\bibinfo{volume}{830}}, \bibinfo{pages}{51} 
 (\bibinfo{year}{2016}). 


\bibitem{Stefanon2021} 
 \bibinfo{author}{Stefanon, Mauro, et al.}  
 \newblock \bibinfo{title}{{The Spitzer/IRAC Legacy over the GOODS Fields: Full-depth 3.6, 4.5, 5.8, and 8.0 $\mu$m Mosaics and Photometry for $>$9000 Galaxies at $z\sim3.5-10$ from the GOODS Reionization Era Wide-area Treasury from Spitzer (GREATS)}}. 
 \newblock \emph{\bibinfo{journal}{The Astrophysical Journal Supplement Series}} 
 \textbf{\bibinfo{volume}{257}}, \bibinfo{pages}{68} 
 (\bibinfo{year}{2021}). 

\end{thebibliography}
\end{document}